\documentclass[conference]{IEEEtran}
\pdfoutput=1
\usepackage{mathptmx} 

\usepackage[normalem]{ulem}
\usepackage[hyphens]{url}
\usepackage[sort,nocompress]{cite}
\usepackage[final]{microtype}
\usepackage[keeplastbox]{flushend}

\usepackage{amsmath,amssymb,amsfonts}      
\usepackage{algorithmic}
\usepackage{graphicx}
\usepackage{textcomp}
\usepackage{xcolor}
\usepackage{fancyhdr}
\usepackage[hyphens]{url}
\usepackage[normalem]{ulem}
\usepackage{caption}
\usepackage{subcaption}
\usepackage{graphics}
\usepackage{float}
\usepackage{dblfloatfix}   
\usepackage{multirow}
\usepackage{xspace}
\usepackage{comment}
\usepackage{xcolor,colortbl}
\usepackage{tikz}

\usepackage{soul}
\usepackage[bookmarks=true,breaklinks=true,letterpaper=true,colorlinks,linkcolor=black,citecolor=blue,urlcolor=black]{hyperref}
\usepackage{enumitem}
\usepackage[ruled,vlined]{algorithm2e}
\usepackage{listings}

\usepackage{adjustbox}

\setlist{nosep}
\setlist{noitemsep}
\definecolor{Gray}{gray}{0.85}

\newcommand{\SYMBOL}{\mathit{symbol}}
\newcommand{\OFFSET}{\mathit{offset}}

\newcommand{\OURLCORE}{APack} 

\newcommand{\OURL}{\textit{\OURLCORE}\xspace} 

\def\BibTeX{{\rm B\kern-.05em{\sc i\kern-.025em b}\kern-.08em
    T\kern-.1667em\lower.7ex\hbox{E}\kern-.125emX}}

\title{APack: Off-Chip, Lossless Data Compression for Efficient Deep Learning Inference}
\author{\large 
Alberto Delmas Lascorz\textsuperscript{1} \ \ \ \ \ \ \ \ 
Mostafa Mahmoud\textsuperscript{1} \ \ \ \ \ \ \ \ Andreas Moshovos\textsuperscript{1,2}\\
\small
\textsf{a.delmaslascorz@mail.utoronto.ca \ \ mostafa.mahmoud@mail.utoronto.ca \ \ moshovos@eecg.toronto.edu} \\
\large
\textsuperscript{1}University of Toronto\\
\textsuperscript{2}Vector Institute}

\begin{document}
\maketitle
\thispagestyle{plain}
\pagestyle{plain}


\begin{abstract}

Data accesses between on- and off-chip memories account for a large fraction of overall energy consumption during inference with deep learning networks. We present \OURL, a simple and effective, lossless, off-chip memory compression technique for fixed-point quantized models. \OURL reduces data widths by exploiting the non-uniform value distribution in deep learning applications. \OURL can be used to increase the effective memory capacity, to reduce off-chip traffic, and/or to achieve the desired performance/energy targets while using smaller off-chip memories. \OURL builds upon arithmetic coding, encoding each value as an arithmetically coded variable length prefix, plus an offset. To maximize compression ratio a heuristic software algorithm partitions the value space into groups each sharing a common prefix. \OURL exploits the memory access parallelism of machine learning workloads to replicate and operate several encoder/decoder units in parallel. Combined with the ability to pipeline these units so that they can be time multiplexed across several data streams, allows \OURL to keep up with the data bandwidth demands of the target workloads. In the demonstrated configuration, \OURL is placed just before the off-chip memory controller, where it transparently to the rest of the on-chip system compresses and decompresses data. The rest of the on-chip memory and compute units thus see the original data stream. As a result, \OURL can be used with any machine learning accelerator such as for example vector-like or tensorcore units which are found in modern graphics processors~\cite{durant_2017,DBLP:journals/corr/abs-1804-06826}, systolic arrays such as those used in the Tensor Processing Unit~\cite{TPUISCA17}, and units that process sparse tensors such as those used in the SCNN accelerator~\cite{SCNN}. We implemented the \OURL compressor and decompressor in Verilog and in a 65nm tech node demonstrating its performance and energy efficiency. Indicatively, \OURL reduces data footprint of weights and activations to 60\% and 48\% respectively on average over a wide set of 8-bit quantized models. It naturally adapts and compresses models that use even more aggressive quantization methods. When integrated with a Tensorcore-based accelerator, \OURL boosts the speedup and energy efficiency to $1.44\times$ and $1.37\times$ respectively. 
 \end{abstract}

\section{Introduction}
Much of the overall energy consumption and latency during deep learning inference is due to memory accesses~\cite{energyproblem,Exascale}.  Lossless value compression, by reading and writing fewer bits per value, improves energy efficiency and execution time performance without affecting model accuracy. Since compression relies on value content, it complements other optimization techniques such as dataflow and blocking,~\cite{Eyeriss_reuse,tvm} and quantization~\cite{quantizationsurvey,han_deep_2015-1,choi2018pact,outlierISCA,nikoli2020bitpruning}. The benefits with compression can be particularly pronounced for off-chip accesses as compared to on-chip accesses they incur an order of magnitude more energy and latency~\cite{Horowitz:Energy}.

Memory compression has been proven effective for general purpose applications, however, Delmas et al., demonstrated that methods for general purpose system are not well suited for capturing the unique properties of the value stream of deep learning workloads~\cite{Lascorz:2019:SEF:3352460.3358295}. This has led to the development of specialized compression methods~\cite{han_deep_2015-1,Lascorz:2019:SEF:3352460.3358295,DBLP:conf/hpca/RhuOCPKK18,cavigelli2018bitPlaneCompr,BOVEDA}. Several designs incorporate run-length encoding for zeros~\cite{isscc_2016_chen_eyeriss, EIEISCA16, CambriconISCA16}. Deep Compression quantizes the parameters (weights) into a 16-entry dictionary of 16-bit values which are Huffman encoded off-chip~\cite{han_deep_2015-1}. It uses fine-tuning to recover any accuracy losses. Rhu et al., capitalize on the high frequency of  zeros~\cite{DBLP:conf/hpca/RhuOCPKK18}, Delmas et al., target prefixes of 0s and 1s at the bit-level~\cite{Lascorz:2019:SEF:3352460.3358295}, and Cavigelli et al., pack a group of values bit-level-wise before compression~\cite{cavigelli2018bitPlaneCompr}. Finally, Edo et al. target prefixes of 0s or 1s in the on-chip memory hierarchy~\cite{BOVEDA}. As effective they are, these methods either place additional constraints on the values of the weights while not compressing activations, or target specific value patterns. 

We improve upon past methods by presenting \OURL, a \textit{lossless} \textit{off-chip } compression method that naturally exploits \textit{any} non-uniformity in the distribution of fixed-point weights and activations during inference. \OURL builds upon Arithmetic Coding (AC)~\cite{abramson_1963,Rissanen76,AC1979}, an encoding/decoding method that can achieve nearly \textit{optimal} coding efficiency (e.g., it can encode extremely frequent values using just a \textit{fraction} of a bit). Since its inception, much of work on AC moved along two general directions: Firstly, works that maximize compression ratio such as dynamic adaptation~\cite{CABAC} which generally increase implementation cost and complexity. Secondly, works improving implementation efficiency, e.g.,~\cite{howard_vitter_1992, HEVC}. Regardless, using AC for \textit{off-chip} compression during deep learning inference is challenging for three reasons: 1)~AC implementations remain costly in area, latency, and energy.  2)~Maximizing compression ratio requires assigning a probability of occurrence to each possible data value which in turn needs comparatively large lookup tables which further exacerbates area, latency, and energy costs. 3)~AC is sequential in nature which is at odds with the wide, and high-bandwidth needs of deep learning. 

 \begin{figure} 
\centering
\subfloat[Baseline Accelerator]{\includegraphics[width=.5\linewidth]{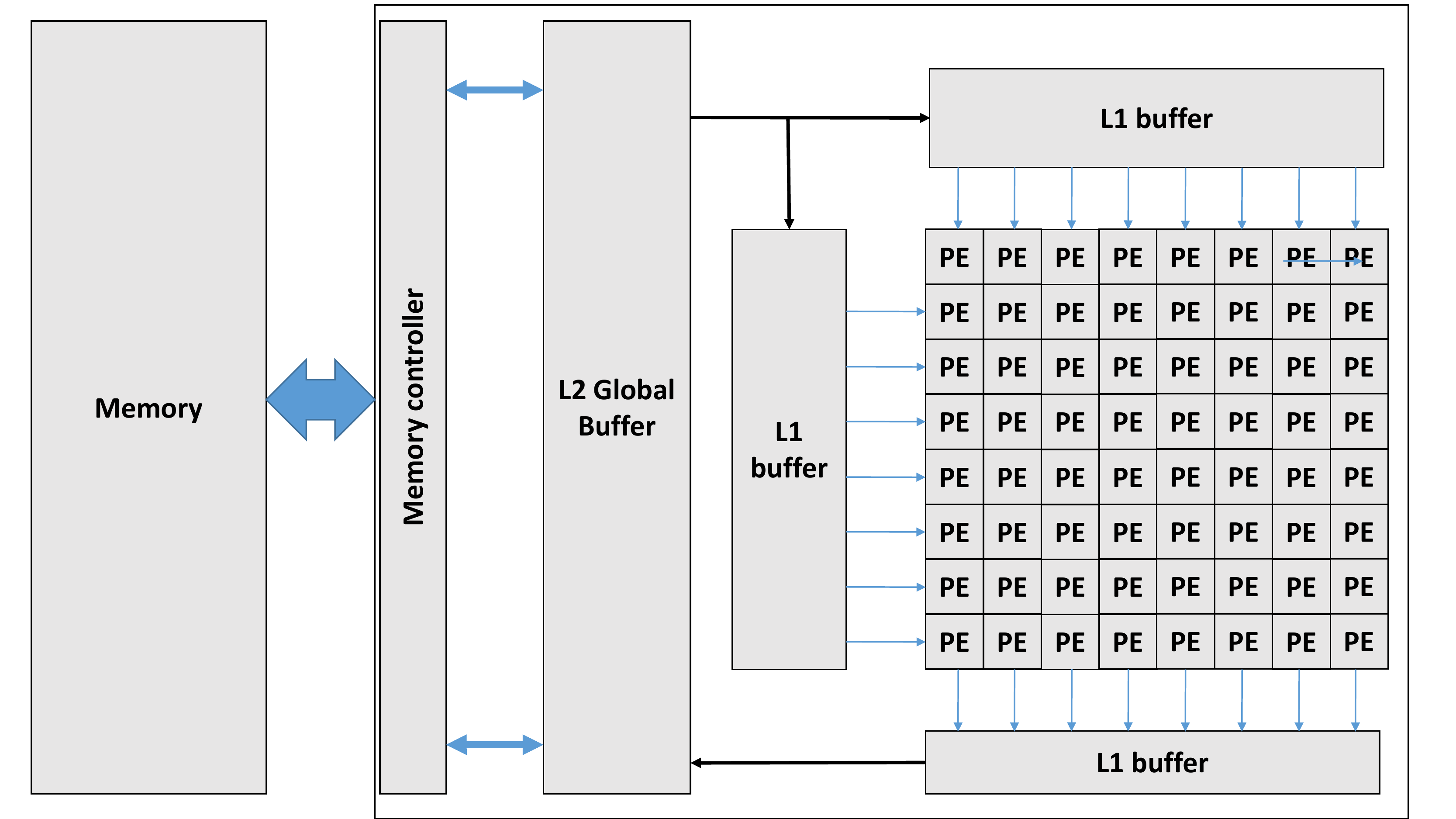}
\label{fig:concept:base}
}
\subfloat[\OURLCORE\ Additions]{\includegraphics[width=.5\linewidth]{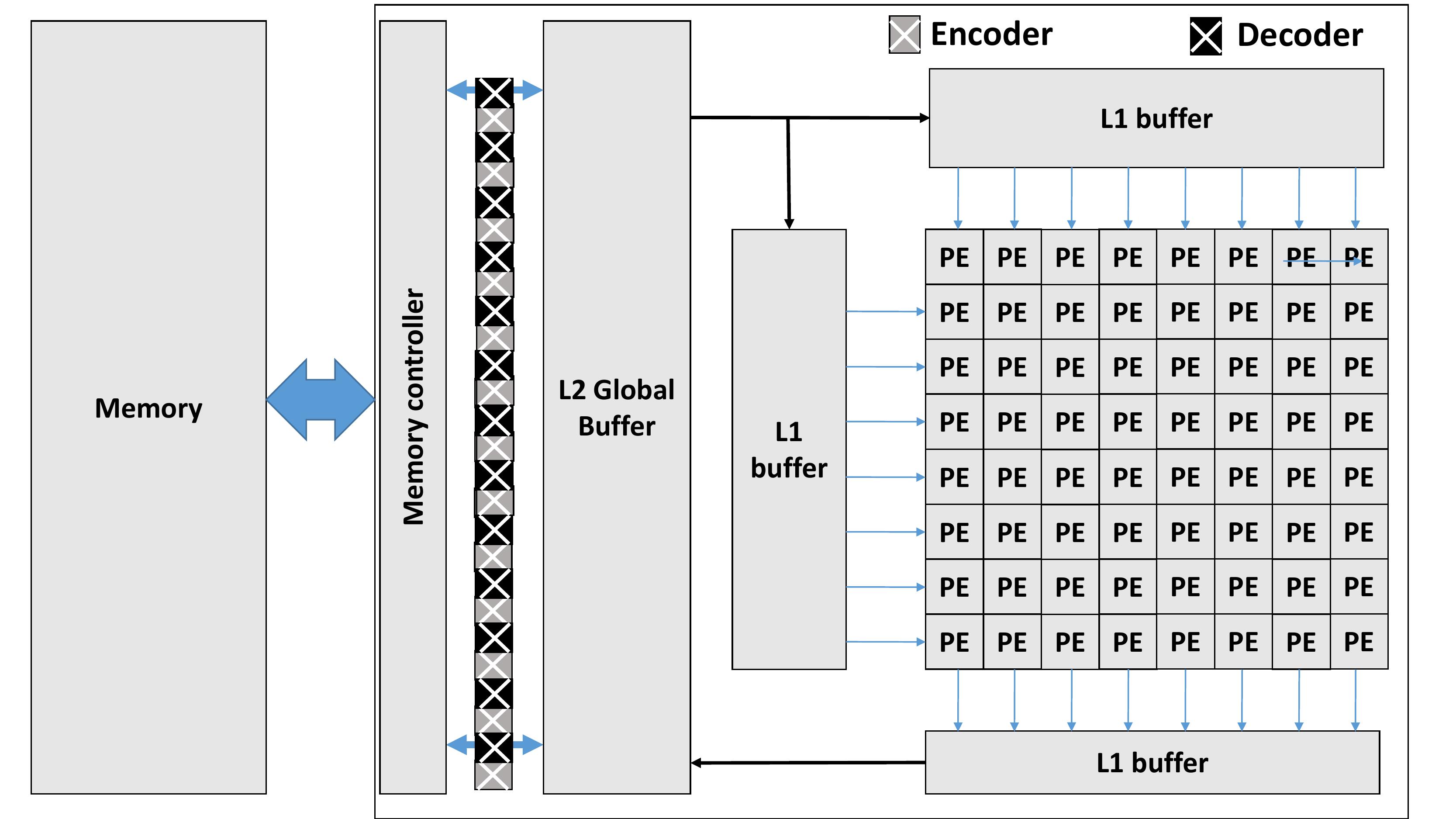}
\label{fig:concept:ours}
}
\vspace{-4pt}
\caption{Incorporating \OURL into an accelerator.}
\label{fig:concept}
\vspace{-10pt}
\end{figure}


To benefit from the high coding efficiency of AC, \OURL balances compression effectiveness vs. implementation cost, energy efficiency, and throughput. For this purpose, \OURL partitions the value space of the activations and weights into sub-ranges so that every value $v$ can be mapped to a $(\SYMBOL, \OFFSET)$ pair such that $ v = \SYMBOL + \OFFSET$ and where the values within the same sub-range share the same prefix $\SYMBOL$. \OURL selects these sub-ranges to maximize encoding efficiency. Intuitively, the more frequent a value is the smaller the sub-range \OURL selects to contain it. At the extreme, a sub-range could even contain just a single frequent value necessitating no offset. \OURL tailors the sub-range selection to the expected value distribution per tensor to maximize coding efficiency. The process is outlined below.

\OURL uses these sub-ranges to encode each tensor into two streams, one encoding the $\SYMBOL$ sequence, and another containing the corresponding offsets. Both streams are read and written sequentially which improves off-chip memory bandwidth utilization and energy efficiency. \OURL uses arithmetic coding only for the $\SYMBOL$ stream. The required probability count and symbol assignment tables (see Section~\ref{sec:Apack}), or \textit{probability tables} for short, \OURL derives using profiling. Each probability count represents the expected frequency of occurrence of a sub-range of values, a key design choice that achieves low cost implementation with little loss in compression effectiveness. A table with just 16 symbols proves sufficient. A heuristic search algorithm constructs this table using a sample run of the network. While individual activations vary considerably with the input, their \textit{overall} distribution per layer varies little enabling \OURL's profile-based approach.  

\OURL is directly compatible with \textit{any} accelerator, such as graphics processors~\cite{durant_2017,DBLP:journals/corr/abs-1804-06826}, systolic arrays~\cite{TPUISCA17}, or grid-like designs targeting sparsity~\cite{SCNN}. Figure~\ref{fig:concept} shows that for an example systolic-array accelerator, \OURL's purpose-built, efficient, and low-cost hardware encoder/decoders sit in-between the on-chip and the off-chip memory hierarchy where they seamlessly decode and encode values as they are being read or written from off-chip. Several encoder/decoder units operating concurrently meet the high \textit{bandwidth} demands of on-chip processing. The off-chip memory hierarchy still sees regular streams of DRAM-friendly wide accesses, albeit ones containing fewer accesses whereas the \textit{on-chip} memory hierarchy and the processing units still see the original values. \OURL works with any desired dataflow transparently packing data in memory so that they can be fed into the corresponding units with little cost. 

Prior to loading the inputs for a layer, \OURL loads the probability count tables into the hardware decoders and then proceeds to decode the input data using sequential accesses to the $\SYMBOL$s and $\OFFSET$ streams. Each decoder produces one value per cycle. To sustain the bandwidth needed by the workloads, \OURL deploys several decoders operating in parallel. For this purpose, the input data stream is partitioned into separate substreams (an approach commonly used in deep learning accelerators~\cite{SamsungISCA2021}), each of which is decoded independently using several decoding units operating concurrently. On the output of each layer one or more encoder units encode the values into $\SYMBOL$ and $\OFFSET$ streams prior to writing to memory. Each encoder can encode one value per cycle. Each encoder and decoder can be pipelined and it can time-multiplex over multiple streams. \OURL's encoders and decoders  use fixed-point logic and two small tables each containing 16 rows of 10b and 11b values respectively. Encoding each stream requires 3 16b registers and 2 8b registers for maintaining state while requires 3 16b registers and 1 8b register. 

\OURL boosts the effective off-chip capacity and bandwidth without requiring any modifications to the neural network model. To a neural network developer \OURL presents a system that needs to go off-chip less often and that rewards well established model optimizations such quantization and pruning without requiring them. More generally, \OURL will reward any method that yields more biased value distributions. 
For system designers \OURL reduces the amount of off-chip memory and thus the cost needed to meet a desired performance target.

We highlight the following experimental results:
\begin{itemize}
\item The \OURL compressor and decompressor when implemented in a commercial 65nm tech node occupy respective $0.02mm^2$ and $0.017mm^2$ area and consume $2.8mW$ and $2.65mW$ of power.
\item The compression rate for the activations of the various DNNs we experimented with vary from $1.43\times$ to $4.2\times$ and is $2.2\times$ on average.
\item The compression rate for the weights vary from $1.13\times$ to $11.4\times$ and is $1.8\times$ on average.
\item \OURL naturally rewards quantization delivering further reductions in traffic as non-uniformity in the value distribution is still present even when extreme quantization is used. It excels at exploiting sparsity for pruned models.
\item When integrated with a Tensorcore-based inference accelerator, \OURL unlocks a speedup of $1.44\times$ by avoiding stalls for off-chip transfers and enabling much better use of the accelerator’s compute units. In addition, \OURL boosts energy efficiency of the accelerator by $1.37\times$.
\end{itemize}

\section{Challenges and Opportunities}
\label{sec:dists}

\begin{figure}[t!] 
\centering
\includegraphics[width=.9\linewidth]{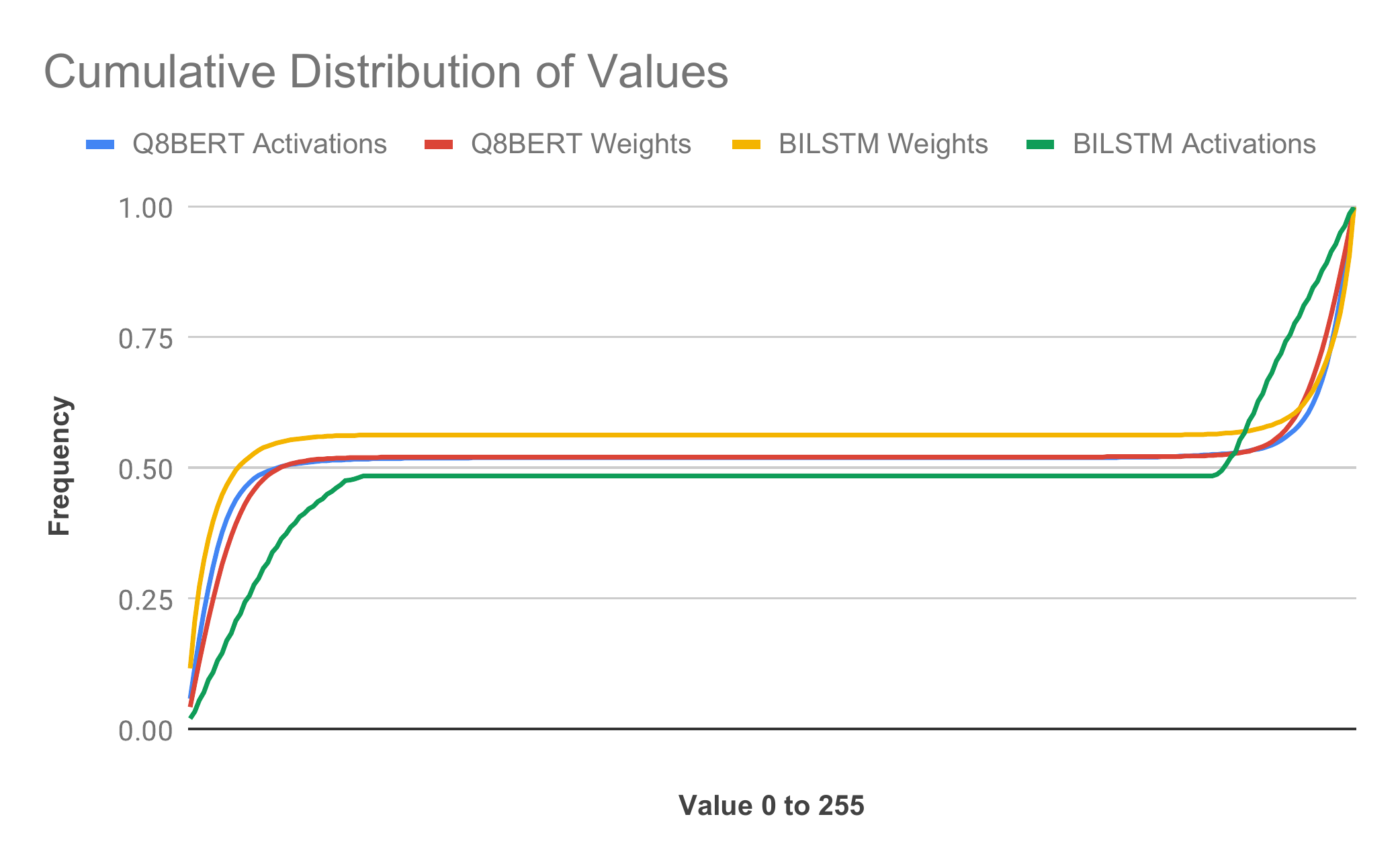}
\caption{Cumulative distribution of weight and activation values for layers from Q8BERT and BILSTM.}
\label{fig:motivation}
\vspace{-10pt}
\end{figure}

As motivation Figure~\ref{fig:motivation} presents the cumulative distribution of the frequency of occurrence for the values of two layers. The first is Layer 10 from Q8BERT and the second is Layer 1 from BILSTM (see Table~\ref{tbl:models} for a list of all the models studied) both quantized to 8b fixed-point. The distributions are far from uniform: Around half of the values tend to be close to zero, where another half or so tends to be close to 255. Conventional memory hierarchies do not capitalize on this property as they store all values using a container which is sufficiently long to accommodate \textit{any} value possible. Prior compression work takes advantage of such distributions(see Section~\ref{sec:related}). For example, Shapeshifter does not store prefixes of 0s (group near zero) or 1s (group near 255 . However, as the bit length is encoded explicitly using an extra field, the values are first grouped so that they can all use the same size container. This reduces encoding efficiency. It also uses a bit vector to remove zero values which further reduces efficiency when zeros are not as frequent.

In contrast, arithmetic coding does not rely on value magnitude or any in value content in general. Instead, it utilizes a number of bits that is proportional to the \textit{frequency} of each value. It outperforms other encoding methods that rely on frequency such as Huffman coding, as it can assign even a fractional number of bits to each input value. 


\section{Arithmetic Coding Primer}

This section is a refresher of arithmetic coding fundamentals -- more appropriate introductions are available~\cite{ACintro, Nelson:AC:DrJobbs}. At a high-level arithmetic coding converts an input sequence of symbols (in our case values) into a \textit{code} which is a value between 0 and 1. The code uniquely identifies the input sequence and can be used to reproduce it. The precision, i.e., the number of bits needed to represent this code depends on the frequency of each value. For maximal compression efficiency, arithmetic coding needs arbitrary precision number precision and arithmetic.

Let us consider an example of encoding four values (symbols in AC parlance) A through D, respectively with frequencies of 0.4, 0.1, 0.3, 0.2. Arithmetic coding could work by assigning the range $[0,0.4)$ to A, the range $[0.4, 0.5)$ to B, and $[0.5,0.8)$ and $[0.8,1.0)$ respectively to C and D. Then a \textit{single} A can be represented by any number in the $[0,0.4)$ range, whereas any value in $[0.4,0.5)$ represents a single B. While this may seem inefficient, this is only because we are looking presently at single values. Let us consider instead the sequence ABAC. Arithmetic coding progresses by maintaining a range $[low,high)$ which initially is $[0.0,1.0)$. It first encodes A restricting the range to $[0.0,0.5)$. To encode the B that follows, it further restricts the current range to its $[0.4,0.5)$ sub-range. The new range becomes $[0.20,0.25)$. To encode the next A, the coder further restricts the current range to its $[0.0,0.5)$ sub-range. Thus the range becomes $[0.2000,0.2025)$. Notice that prior to A, the range had a width of $0.05 = 0.25 - 0.20$ and that the high mark for A is $0.5$ making the new high mark $0.20 + 0.05 \times 0.5$ off the new low (which stays at $0$ as A is assigned the range starting at $0$).  Encoding the final B requires further restricting the existing range to its $[0.8,1.0)$ sub-range, or to $[0.2002,0.2025)$. Any value within this range can be used to represent the encoded sequence of ABAC. The longer the input sequence, the more precision the resulting encoded number will need. It is for this reason that as presented AC requires arbitrary precision arithmetic.

Formally, Arithmetic Coding accepts a sequence $S$ of input symbols $S=s_N...s_i...s_0$ from vocabulary $V$ of symbols $\{v_s,...,v_0\} \in V$ and a table of ranges $[phigh_j, plow_j)$, one per symbol in $V$. For maximal compression, the size of each range has to be proportional to the probability of occurrence of the respective symbol. Arithmetic coding outputs a $code$ value, a number in $[0,1)$ which uniquely represents the input sequence $S$. Internally, the method uses two state variables $high$ and $low$. The following is a pseudo-code implementation:

\begin{algorithm}
\SetAlgoLined
\LinesNumbered
\footnotesize
\KwData{$S=s_N,...,s_0$, $phigh_s,...,phigh_0$, $plow_s,...,plow_0$}
\KwResult{$code$ representing the input sequence $S$}
    $low = 0.0;$ $ high =1.0$;
    
    \While{ i $<$ N}{
        $s = s_i$ 
        
        $probh = phigh_{s_i}$; $probl = plow_{s_i}$   
        
        $range = high - low + 1$ 
        
        $high = low + probh / range$  
        $low = low + probl / range$
        
        i++
    }
    $code \gets low$
\caption{Basic Arithmetic Coding: Infinite precision method.}
\end{algorithm}
 
Encoding starts by setting the current range boundaries to $[0.0,1.0)$ in line 1.
Each symbol $s_i$ is read in line 3, and it is used to index the table of ranges in line 4. Line 3 calculates the current range length $range$. The new boundaries are offset from the current low boundary, by adding the scaled with $range$ symbol boundaries. 

\section{\OURLCORE}
\label{sec:Apack}
While effective, the presented arithmetic coding method has features that are undesirable for hardware implementation: 1)~It requires infinite precision arithmetic, and 2)~uses a range table with one entry per potential input value.

Approximations of infinite precision arithmetic may be possible, however, they are expensive, moreover they would require execution time that is proportional to the current precision. As we expect to be encoding/decoding tensors of several thousands of elements, even if the implementation cost was not prohibitive, the execution time, and energy would render this approach unprofitable. Fortunately, there are implementations of arithmetic encoding that use finite precision. \OURL's arithmetic coding is inspired by Nelson's software implementation~\cite{Nelson:AC:DrJobbs}. \OURL however, implements a single step encoding/decoding process where all updates to the state (high, low, and code) are done in a single step (the aforementioned implementation updates and produces one bit at a time) and where arithmetic coding is used only for a variable portion of each input value (the aforementioned arithmetic coding assumes that whole values of a fixed-length are encoded). 

Using a table with one entry per potential input value would be prohibitively expensive energy- and area-wise for our purposes. Given that the encoder and decoder process one symbol (values) per step, we would like to keep area and energy costs to a minimum so that we can replicate the units to achieve high bandwidth data supply. For 8b models we would need to have at least one table of $256 \times 10b \times 2  \approx 5Kb$ of storage just for the range table (the overheads would be even higher for 16b models which are still used in certain applications that require high resolution output such as segmentation). Even if the area was not a concern, the energy required to access such a table would be prohibitive. 

Instead of assigning a table entry per symbol, \OURL uses a limited number of entries by  partitioning the input value space into several non-overlapping ranges [$v_{min}$, $v_{max}$]. Every value $v$ within the range is encoded as $(\SYMBOL,\OFFSET)$ where $\SYMBOL = v_{min}$ and $\OFFSET = v - \SYMBOL$ an unsigned integer of $OL=lg(v_{max}-v_{min})$ bits. Additionally, $v_{min}$ is always of the form $x...x0...0$, that is it a bit prefix of all values within the range. We have experimented with several 4b, 8b, and 16b models and found that using 16 ranges, with 8b $v_{min}$ and $v_{max},$ and 3b $OL$ (4b for 16b models) is sufficient.   

The \OURL encoder accepts an input sequence of values and compresses it, a value at a time, into two sequences. It encodes each value into a $(\SYMBOL, \OFFSET)$ pair according to the range it maps to. It then arithmetically encodes just the $\SYMBOL$ ($v_{min})$ while storing the $\OFFSET$ verbatim using only as many bits as necessary (very frequent symbols may end up using no $\OFFSET$ bits). The encoded symbols comprise the first output stream, and the corresponding offsets comprise the second stream. The process completes when the last symbol is encoded. Along with the two encoded streams, \OURL also stores metadata that contain: 1)~the number of symbols encoded (this is used to terminate decoding), and 2)~the range table and the probability table (a total of 298 bytes) used by the arithmetic coder. 

To perform arithmetic coding each value range is also assigned a probability count range $(low_i, high_i)$. We use 10b probability counts, and we always assign the full range of $(0x0, \mathit{0x3ff})$ across the symbols. A software profiler (see Section~\ref{sec:profiler}, uses a heuristic algorithm to determine value ranges that reduce the overall footprint comprising the encoded symbol and offset streams. 

Table~\ref{tbl:symt:ex} shows an example symbol table for the weights of a layer of BILSTM (see Section~\ref{sec:eval}). 
The fields ''IDX'' and 'p' respectively report the row index and the symbol probability and are shown for clarity --- they are not stored. Row 0 captures the four values in the range $[0x00,0x03]$ and associates all with the probability count range of $[0x000, 0x1EB)$ which corresponds to a probability 'p' of $0.4795$. Any value in this range will be mapped to symbol $0$ which during decoding will be mapped back to $v_{min}=0x00$. To recover the original number, an $OL=2$b will be recorded such that $v=v_{min}+\mathit{offset}$. Rows 3 through 12 are all assigned to a zero length probability count range ($[0x23A,0x23A)$) where $low = high$). These are values that do not appear at all in the input weight tensor. Since weights are statically known, this is permissible.
Row 13 captures all values in the range of $[0xD0,0xF3]$ which will be mapped to symbol $13$. Notice that the offset requires $6$ bits since $0xF0-0xD0=0x23$. Since $0x23 < 2^6-1$ this means that not all offset values will be used for this range. No special processing is needed to ensure that this is so. If we implement the table using all fields shown, then entries with 0 probability can be omitted, and the order of rows can be changed at will enabling power gating opportunities. 
In the implementation studied the symbols are ordered such as that $v_{min}[i] = v_{max}[i-1]+1$ for $i>0$ so that we need to store only one of the two per row. Similarly we only store the $high$ count per row.

\begin{table}[]
\caption{Symbol and Probability Count Table Example}
\label{tbl:symt:ex}
\footnotesize
\centering
\ttfamily
\begin{tabular}{|
>{\columncolor[HTML]{C0C0C0}}r |r|l|c|r|r|
>{\columncolor[HTML]{C0C0C0}}r |}
\hline
\textit{\textbf{IDX}} & \textbf{v\_min} & \textbf{v\_max} & \textbf{OL} & \textbf{low} & \textbf{high} & \textit{\textbf{p}} \\ \hline
\textit{0}            & 0x00            & 0x03            & 2           & 0x000        & 0x1EB         & \textit{0.4795}     \\ \hline
\textit{1}            & 0x04            & 0x07            & 2           & 0x1EB        & 0x229         & \textit{0.0605}     \\ \hline
\textit{2}            & 0x08            & 0x0F            & 3           & 0x229        & 0x238         & \textit{0.0146}     \\ \hline
\textit{3}            & 0x10            & 0x3F            & 6           & 0x238        & 0x23A         & \textit{0.0020}     \\ \hline
\textit{4}            & 0x40            & 0x4F            & 4           & 0x23A        & 0x23A         & \textit{0.0000}     \\ \hline
\textit{5}            & 0x50            & 0x5F            & 4           & 0x23A        & 0x23A         & \textit{0.0000}     \\ \hline
\textit{6}            & 0x60            & 0x6F            & 4           & 0x23A        & 0x23A         & \textit{0.0000}     \\ \hline
\textit{7}            & 0x70            & 0x7F            & 4           & 0x23A        & 0x23A         & \textit{0.0000}     \\ \hline
\textit{8}            & 0x80            & 0x8F            & 4           & 0x23A        & 0x23A         & \textit{0.0000}     \\ \hline
\textit{9}            & 0x90            & 0x9F            & 4           & 0x23A        & 0x23A         & \textit{0.0000}     \\ \hline
\textit{10}           & 0xA0            & 0xAF            & 4           & 0x23A        & 0x23A         & \textit{0.0000}     \\ \hline
\textit{11}           & 0xB0            & 0xBF            & 4           & 0x23A        & 0x23A         & \textit{0.0000}     \\ \hline
\textit{12}           & 0xC0            & 0xCF            & 4           & 0x23A        & 0x23A         & \textit{0.0000}     \\ \hline
\textit{13}           & 0xD0            & 0xF3            & 6           & 0x23A        & 0x23C         & \textit{0.0020}     \\ \hline
\textit{14}           & 0xF4            & 0xFB            & 3           & 0x23C        & 0x276         & \textit{0.0566}     \\ \hline
\textit{15}           & 0xFC            & 0xFF            & 2           & 0x276        & 0x3FF         & \textit{0.3838}     \\ \hline
\end{tabular}
\end{table}

The \OURL decoder accepts as input two sequences: 1)~the compressed symbols and 2)~the corresponding offsets. It outputs the original values. At each step, it uses arithmetic decoding on the symbol sequence to obtain first the value prefix that corresponds to the next symbol. Using the symbol table it then extract the appropriate number of offset bits, which it adds to the value prefix. The process continues until all symbols have been decoded. Our decoder produces a single value per step. Using pipelining and replication we can achieve the desired bandwidth target.


\begin{figure*}
\centering
\begin{subfigure}[b]{.25\linewidth}
\centering
\adjustbox{trim={.1\width} {.1\height} {0.1\width} {.0\height},clip}%
  {\includegraphics[width=1.45\linewidth,page=7]{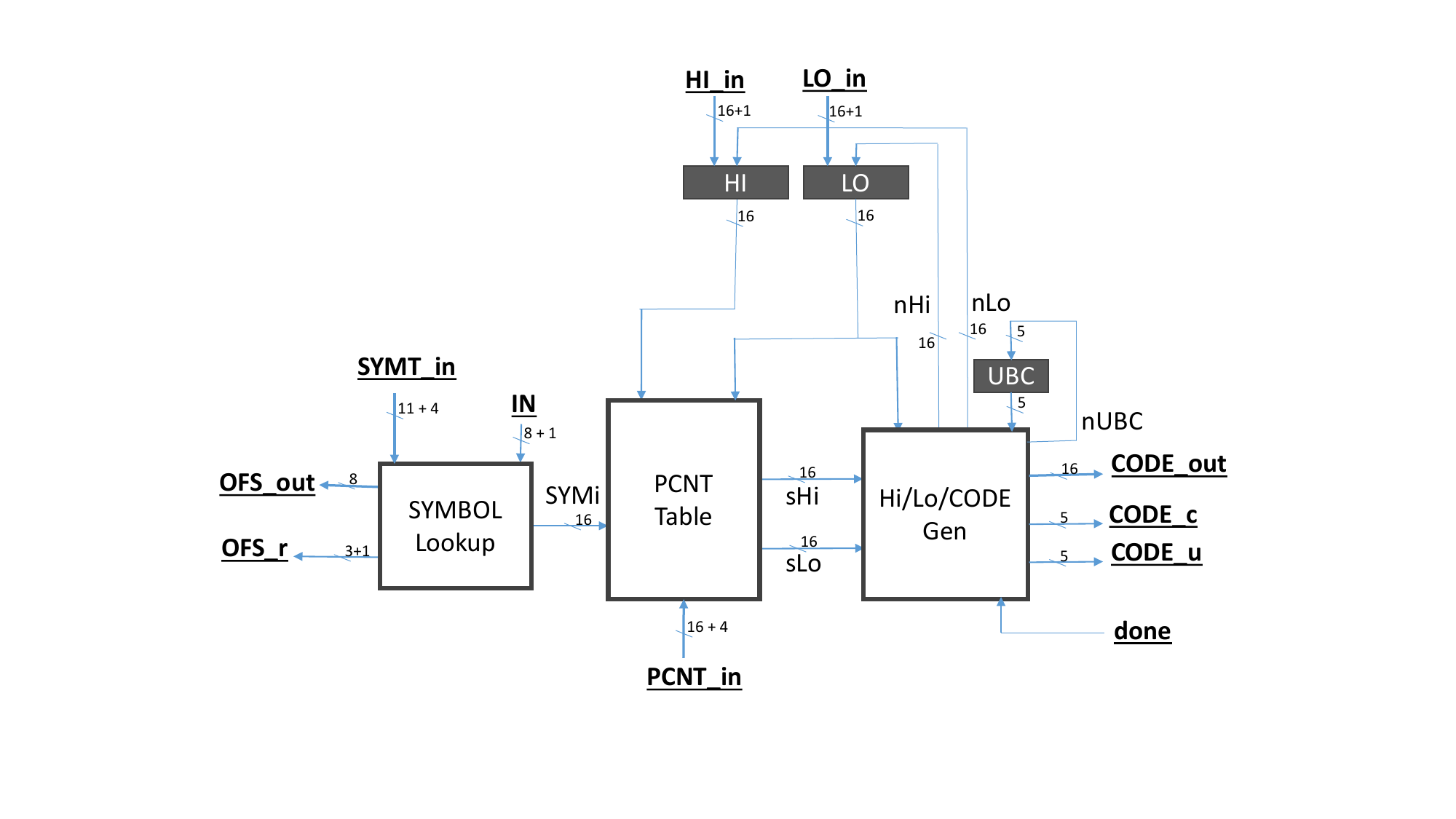}}
\caption{Interface and Organization}
\label{fig:enc:org}
\end{subfigure}
\begin{subfigure}[b]{.28\textwidth}
\centering
\adjustbox{trim={.15\width} {.1\height} {0.1\width} {.0\height},clip}%
  {\includegraphics[width=1.4\linewidth,page=8]{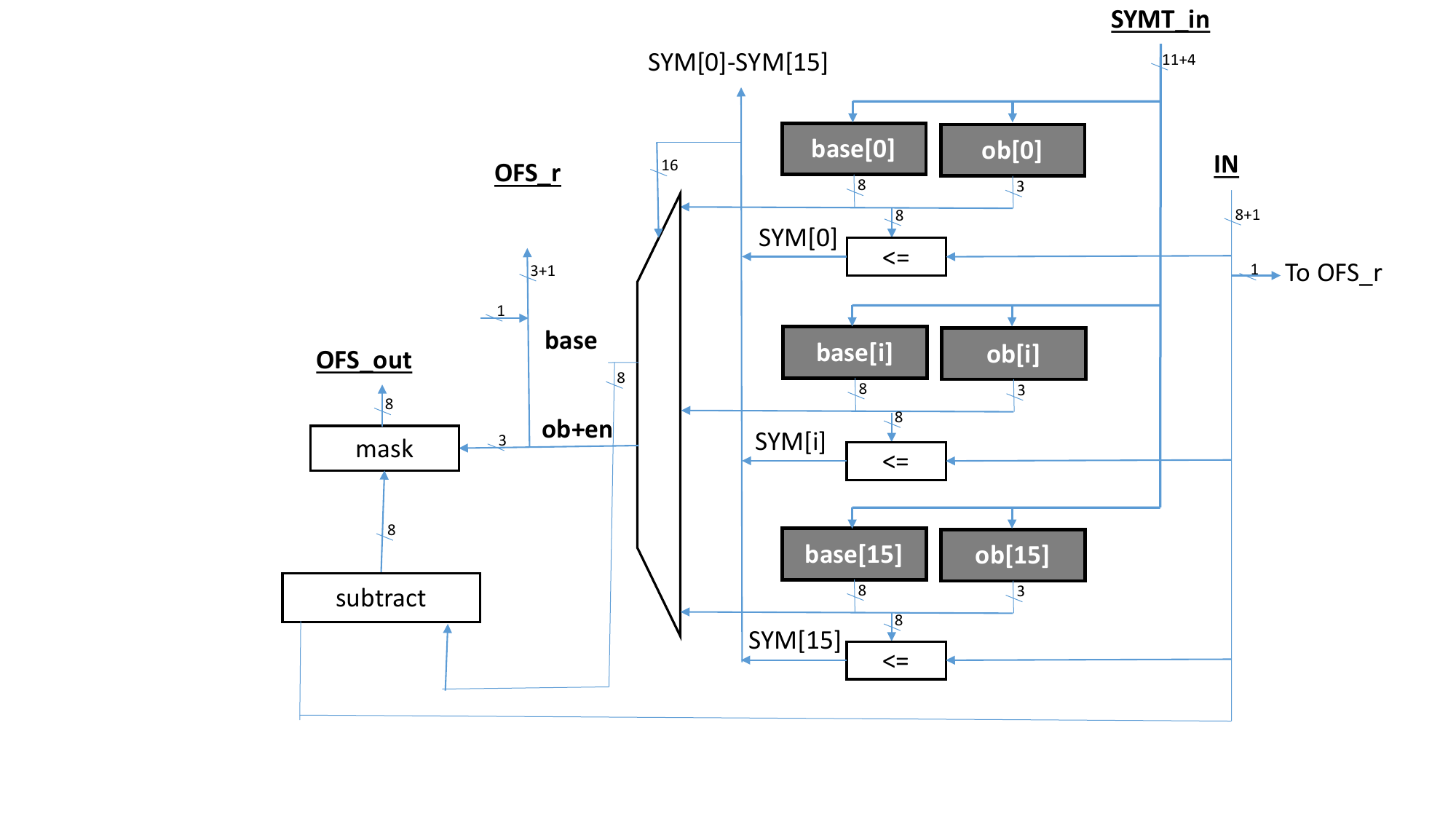}}
\caption{Symbol Lookup.}
\label{fig:enc:symlook}
\end{subfigure}
\begin{subfigure}[b]{.13\textwidth}
\centering
\adjustbox{trim={.10\width} {.05\height} {0.3\width} {.0\height},clip}%
  {\includegraphics[width=2.3\linewidth,page=9]{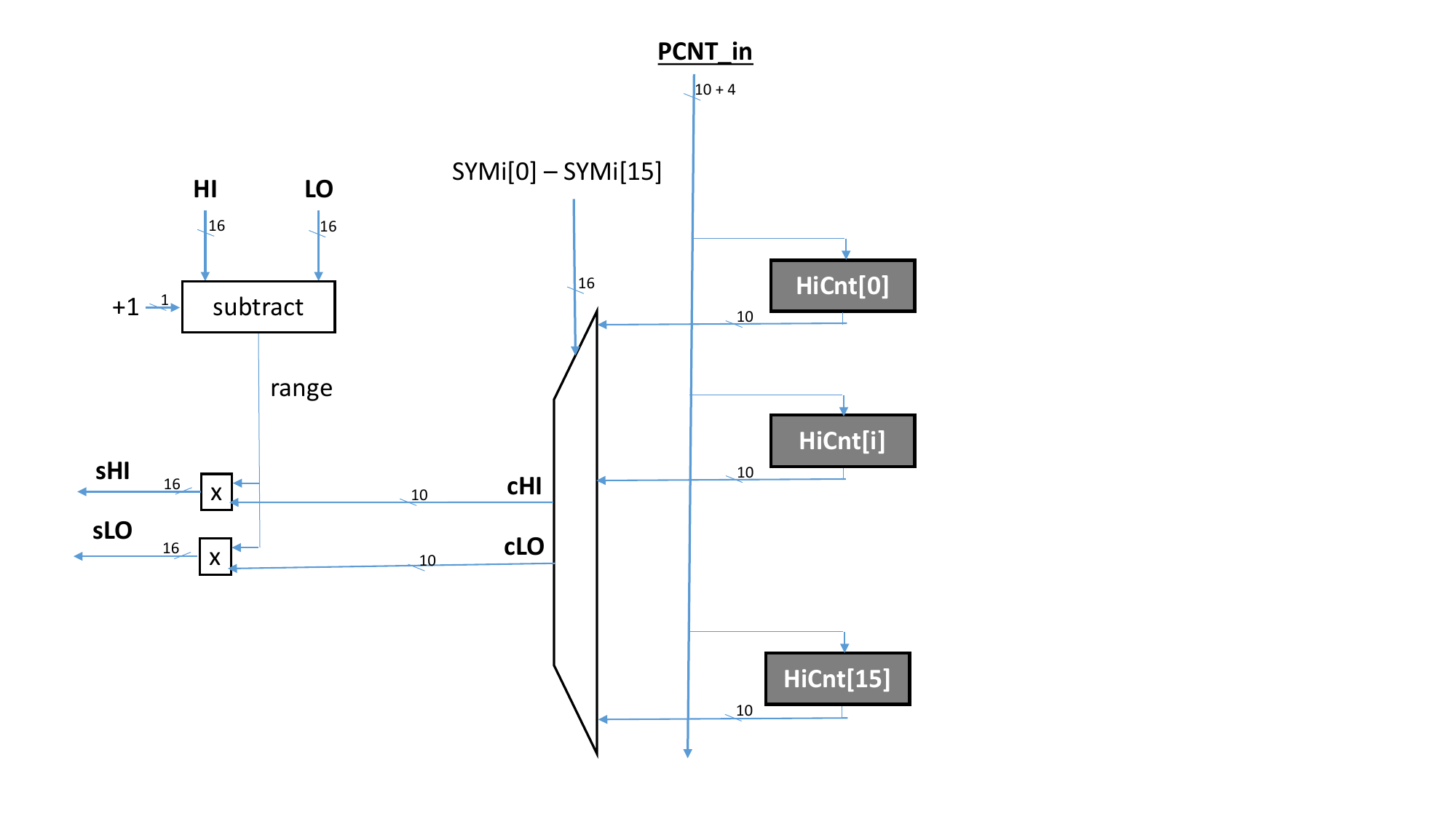}}
\caption{PCNT Table.}
\label{fig:enc:pcnt}
\end{subfigure}
\begin{subfigure}[b]{.29\textwidth}
\centering
\adjustbox{trim={.0\width} {.1\height} {0.0\width} {.0\height},clip}%
  {\includegraphics[width=1.2\linewidth,page=10]{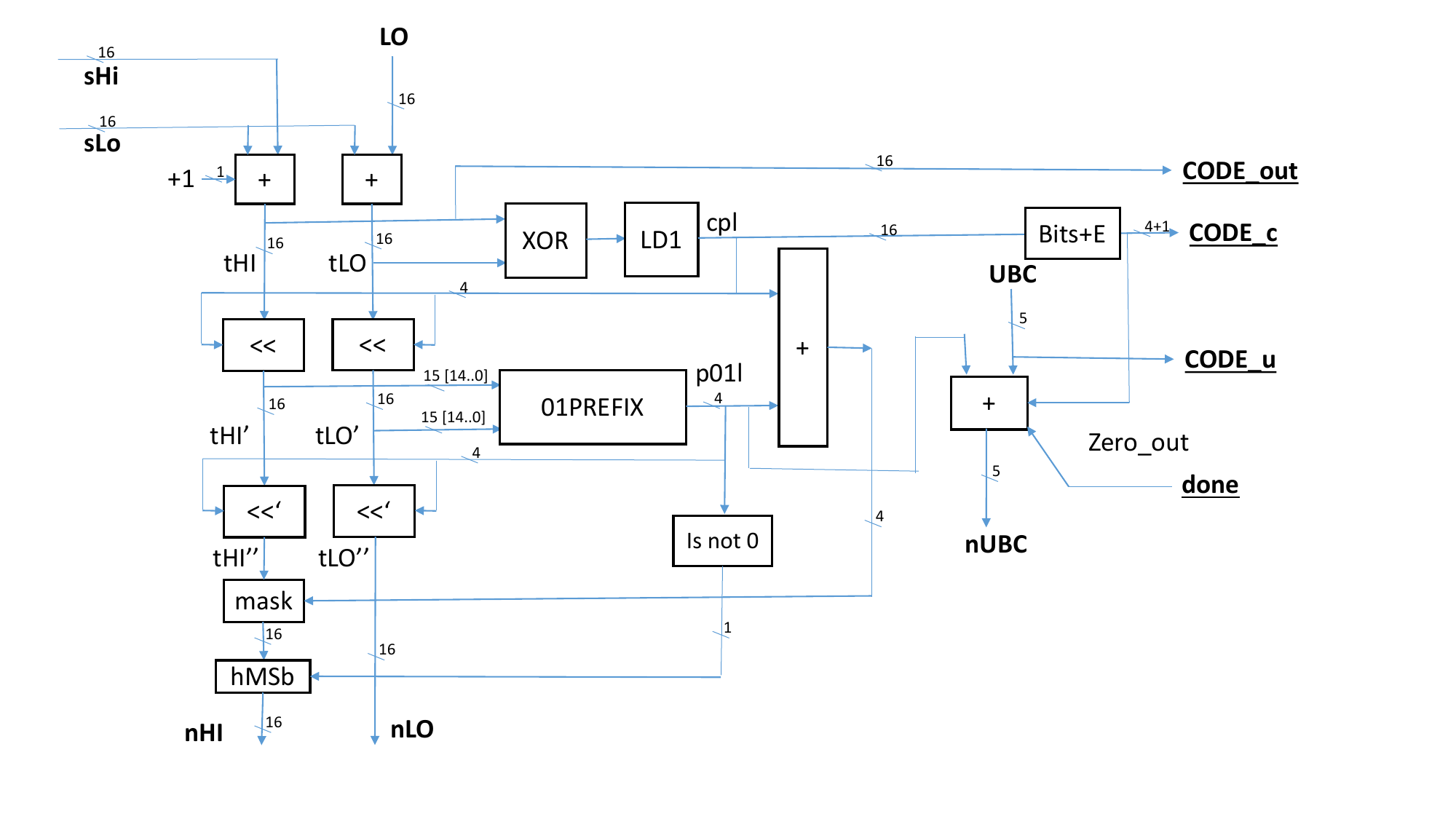}}
\caption{Hi/Lo/Code Generation.}
\label{fig:enc:hilocode}
\end{subfigure}
\caption{\OURLCORE\ Encoder: {Greyed blocks are registers all else is combinatorial logic.}}
\end{figure*}

\section{\OURLCORE~Implementation}
\label{sec:impl:enc}

Figure~\ref{fig:enc:org} shows the interface and high-level organization of an example implementation of the \OURL encoder. The following ports are used for initialization \textit{once} before processing each layer. We use separate encoders for activations and weights: 1)~HI\_in and LO\_in: initialize the decoder range in two internal 16b registers (the extra 1b signal is an enable). 2)~SYMT\_in: initialize the symbol table's value ranges and offsets. The implementation assumes that: i)~the full range is mapped, and ii)~the rows are ordered in value (symbol) order. 3)~PCTN\_in: initialize the probability count entries, each using 10b. It is ordered to match the symbol table.4)~The $8b$ IN port along with its $1b$ enable are used to send the input value that needs to be decoded in this step (if any -- enable). 5)~The $1b$ done signal terminates encoding.

Every step the encoder receives a value to encode via the IN port. 
The encoder processes the symbol internally and produces the following outputs:
1)~OFS\_out: an up to $8b$ value containing the corresponding offset and its 4b length OFS\_r (range 0b-8b).2)~A $16b$ CODE\_out contains CODE\_c (4b+1b to indicate no output) useful bits that are to be written into the encoded symbol stream.  The $5b$ OUT\_u along with its $1b$ enable indicate whether we should include additional ``underflow'' bits (explained below) in the output and how many (count is the $5b$ portion). The bits are to be inserted after the most significant bit of CODE\_out (and they set to its inverse).

The encoder comprises three blocks: ``SYMBOL Lookup'', ``PCNT Table'', and ``Hi/Lo/CODE Gen''. Two 16b registers HI and LO maintain the current range (initialized to 0xFFFF and 0x0000 respectively). Encoding the input value IN starts in ``SYMBOL Lookup'' which identifies the symbol this value maps to. The block produces two outputs: 1)~the output to the offset stream via OUT\_out and OFS\_r. 2)~SYMi an 1-HOT encoded 16b vector indicating the table row (symbol index) the value mapped to. SYMi is fed into ``PCNT Table''  which determines which probability count range this symbol maps to. The output from ``PCNT Table'' is used by ``Hi/Lo/CODE Gen'' block to update the current code range and to produce any encode stream output bits. A $5b$ UBC register counts the number of underflow bits (explained below). Its value is fed into the block and is updated by the unit using UBCn signal.

\noindent\textbf{SYMBOL Lookup: }
Figure~\ref{fig:enc:symlook} shows the structure of ``SYMBOL Lookup''. The $8b$ IN value is compared against the maximum value (base[i] = $R_{max}-1$) of all ranges using one comparator per row. Since the rows are ordered (ascending), the matching row is the last in order whose base is larger than the input value. This is used to generate the SYMi vector which in turns routes the corresponding base and offset length field (ob) to the components that generate the output offset. The offset is calculated as the difference between the $8b$ IN value and the matching base value and is trimmed via the ``mask'' block to $ob$ bits. Processing proceeds to ``PCNT Table''.

\noindent\textbf{PCNT Table: }
\label{sec:enc:pcnt}
As Figure~\ref{fig:enc:pcnt} shows, the SYMi signals select the boundary counts cHI and cLO for the range assigned to the current symbol. For each symbol index, the unit holds a $10b$ probability count representing the maximum for the range (inclusive). This becomes cHI, and the cLO is taken from the register for the preceding row or 0 for row 0. The cHI and cLO values are then scaled (multiplied) with the current $range = $(HI - LO + 1) producing sHI and sLo. Since range is a 16b number and cHI and cLO are 10b numbers sHI and sLO need 26b. Since our max probability count is $2^{10}-1$ we can discard the 10 least significant bits effectively dividing with our max probability count (this converts the counts into probabilities). Accordingly, the multipliers can omit any parts that serve to produce only those 10 least significant bits. The resulting scaled 16b range boundaries, sHI and sLO, are then fed into  ``HI/LO/CODE Gen''.

\noindent\textbf{HI/LO/CODE Gen: }
\label{sec:enc:hilocode}
The final block, shown in Figure~\ref{fig:enc:hilocode}, starts by calculating the new scaled range tHI and tLO given the newly encountered value. The incoming sHI and sLO values correspond are offset starting from 0. Adjusting this offset to LO producing tHi and tLO shifts this range in the appropriate position representing the sequence of previously seen symbols. The remaining logic: 1)~determines if tHI and tLO have a common prefix which can be output, and 2)~detects whether tHI and tLO are in danger of underflow. This implements arbitrary precision arithmetic while using fixed-point units. 
\noindent\textit{Common Prefix Detection: }The XOR block produces a bit vector that identifies any bit positions that are different between tHI and tLO. The LD1 block then detects the leading position that there is a difference. Any preceeding bits can now be shifted out and written into the encoded symbol stream. This results in updated tHI' and tLO'.

\noindent\textit{Underflow Detection and Handling: } The \OURL encoder uses finite precision arithmetic to execute an algorithm that requires arbitrary precision arithmetic. To do so it effectively maintains a window of 16b into what are high and low range boundaries of arbitrary bit precision. The HI and LO registers contain these 16b windows and conceptually, HI and LO have suffixes comprising an ``infinite'' number of 1s and 0s respectively.

The 16b window are allowed to slide to less significant bits by shifting out any prefix bits that can no longer change. To understand why this works, observe that as arithmetic coding encodes one symbol after the other, the HI and LO boundaries shrink. HI always becomes smaller, while LO always grows larger. However, it should always be the case that HI$>$LO since each symbol encoded has a non-zero probability assigned to it. As HI and LO approach they will grow an increasingly longer common prefix. Those are the bits that the encoder can safely ``discard'' by shifting them out of the HI and LO register while writing them on the encoded stream. 

However, there are cases, where depending on the probability range of a new symbol, and the current range, having a window of just 16b is not enough to appropriately scale the range so that HI remains larger than LO. The case is where HI contains a value of the form 100... and LO a value of the form 011... which means that HI and LO are converging around 0.5. To eventually find our whether they will end up being both above 0.5 or below it, requires being able to perform arithmetic with more than 16b. This happens when the range adjustments done are small enough so they need to affect bits that are not yet within the current window and hence are not physically present. The encoder handles such cases preemptively by entering a state where it records how many \textit{underflow} bits are needed allowing the window to slide. Accordingly, the encoder handles this by identifying, starting from the second most significant bit, any prefix of tHI' and tLO' where tLO is all 0s and tHi is all 1s. This subprefix is shifted away from tHI' and tLO'. This results in tHI'' and tLO''. To detect the length of this subprefix, the encoder uses a leading 1 detector for tHI' (ignoring the MSb) and a leading 0 for tLO' (again ignoring the MSb). The subprefix is the most significant position among the two. This is implemented in the 01PREFIX block. Internally, this block uses a 2-input AND gate per bit position, with one input directly connected to tLO' and the other connected after inversion (NOT) to tHI'. The output of those 15 ANDs goes into a leading 0 detector. The leading 0 position is where the subprefix, if any, ends.

This subprefix is removed from tLO' and tHI' producing tLO'' and tHI''. Its length of is added to the UBC register which counts the number of outstanding underflow bits. Those eventually we will be set to the inverse of the most significant bit of HI (1s or 0s if we end up respectively on the upper or lower part of the range below or above 0.5).

\noindent\textbf{Final HI and LO generation:}
After the common prefix and the underflow subprefix have been discarded, we need to adjust the final HI and LO values. First we need to insert a suffix of 1s in HI to make up for the fact that we shifted out several MSbs. Recall, that HI is meant to slide over a number that has a suffix of infinite 1s. In addition, we need to set the MSb of HI to 1 if we entered underflow bit mode. The final output are the nHI and nLO values which are loaded into the HI and LO registers respectively. Other ways of avoiding underflow in the HI and LO registers are possible. For example, the range may be expanded anytime it drops below half the maximum range.

\begin{figure}
\centering
\begin{subfigure}[b]{.3\textwidth}
\centering
\adjustbox{trim={.01\width} {.1\height} {0.05\width} {.0\height},clip}%
  {\includegraphics[width=1.4\linewidth,page=2]{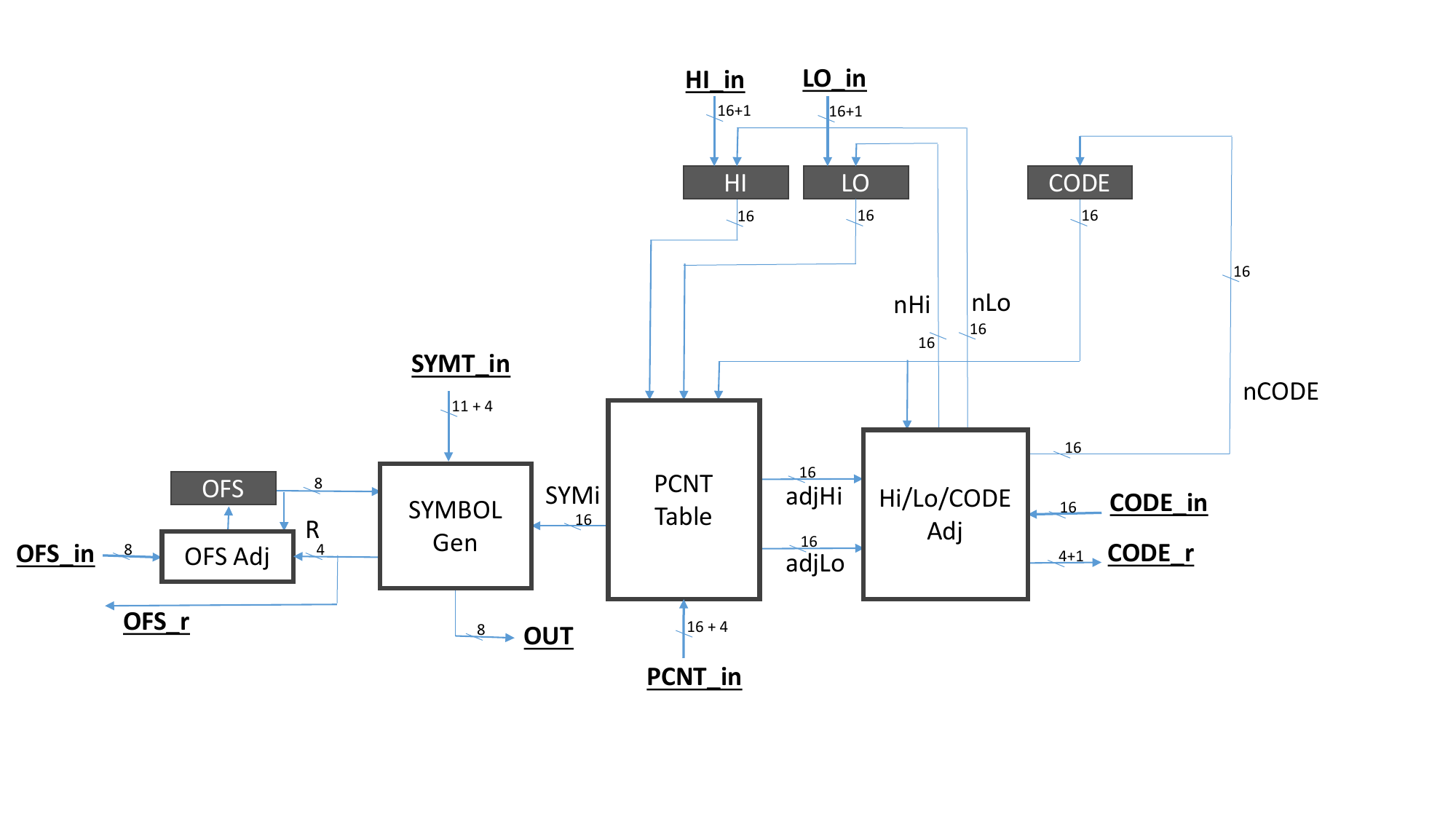}}
\caption{Internal Organization}
\label{fig:dec:org}
\end{subfigure}
\begin{subfigure}[b]{.27\textwidth}
\centering
\adjustbox{trim={.0\width} {.0\height} {0.2\width} {.0\height},clip}%
  {\includegraphics[width=1.1\linewidth,page=3]{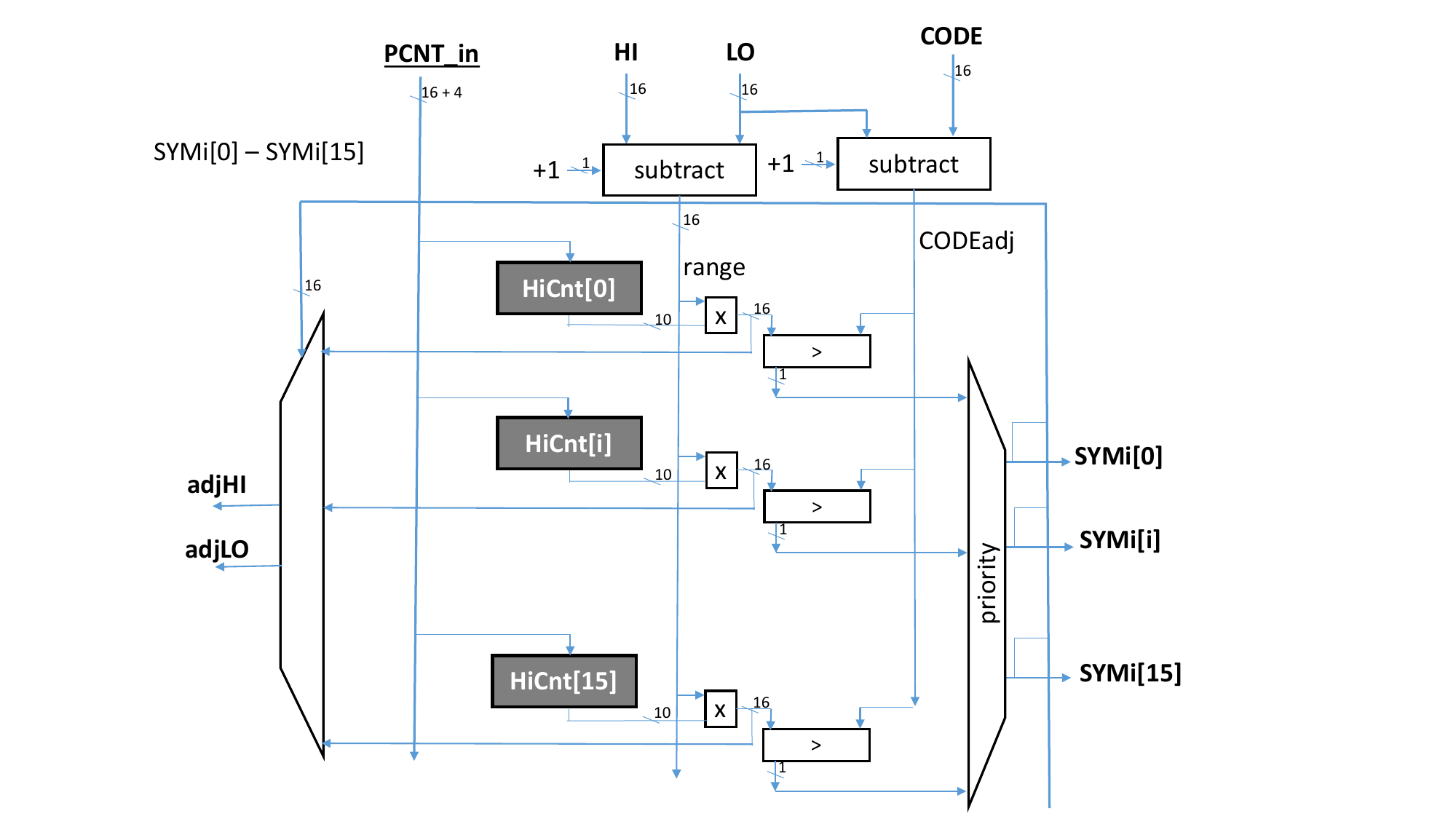}}
\caption{PCNT Table}
\label{fig:dec:pcnt}
\end{subfigure}
\begin{subfigure}[b]{.18\textwidth}
\centering
\adjustbox{trim={.1\width} {.0\height} {0.1\width} {.0\height},clip}%
  {\includegraphics[width=1.6\linewidth,page=5]{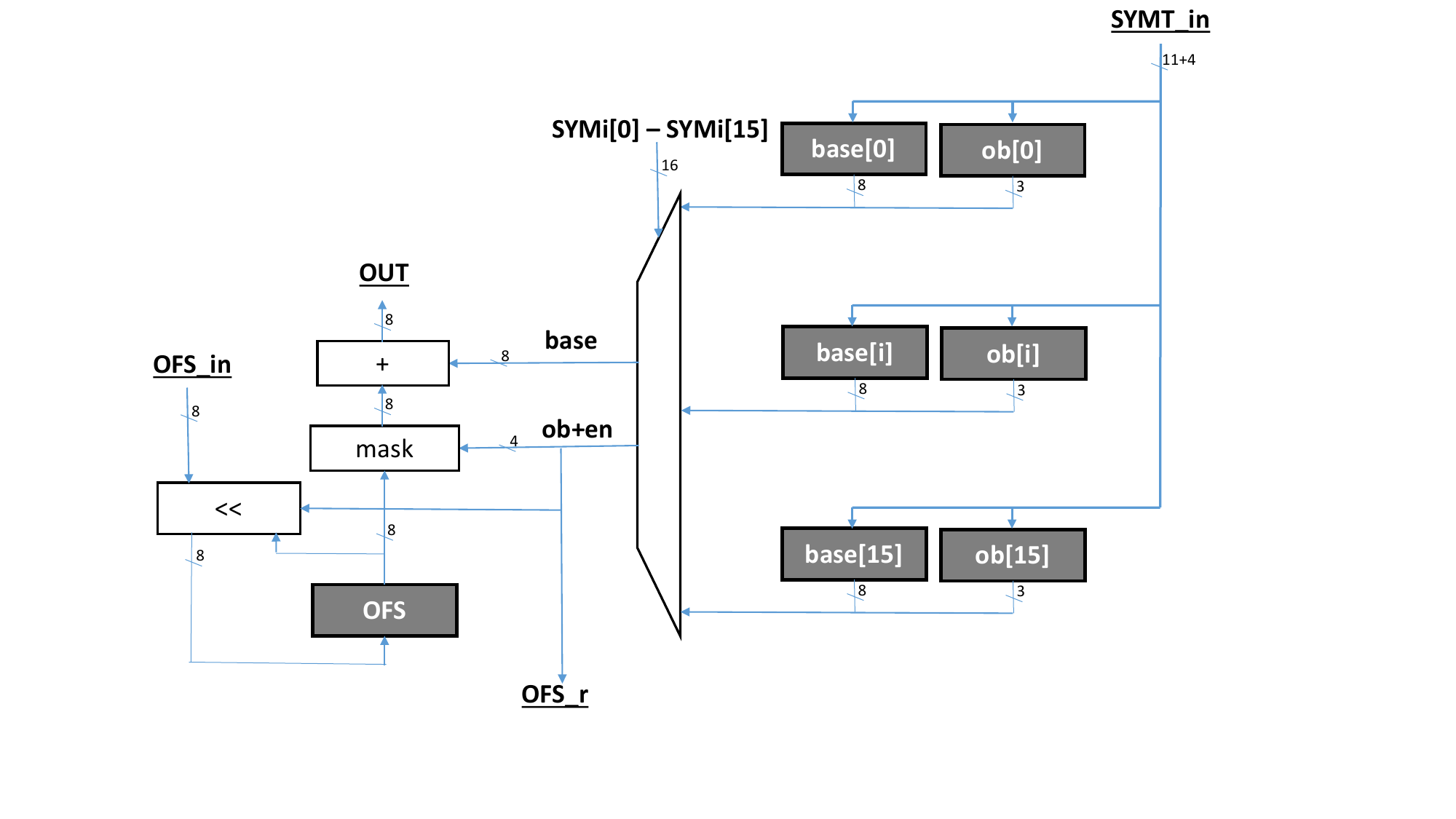}}
\caption{SYMBOL Gen}
\label{fig:dec:symgen}
\end{subfigure}
\begin{subfigure}[b]{.3\textwidth}
\centering
\includegraphics[width=1.3\linewidth,page=4]{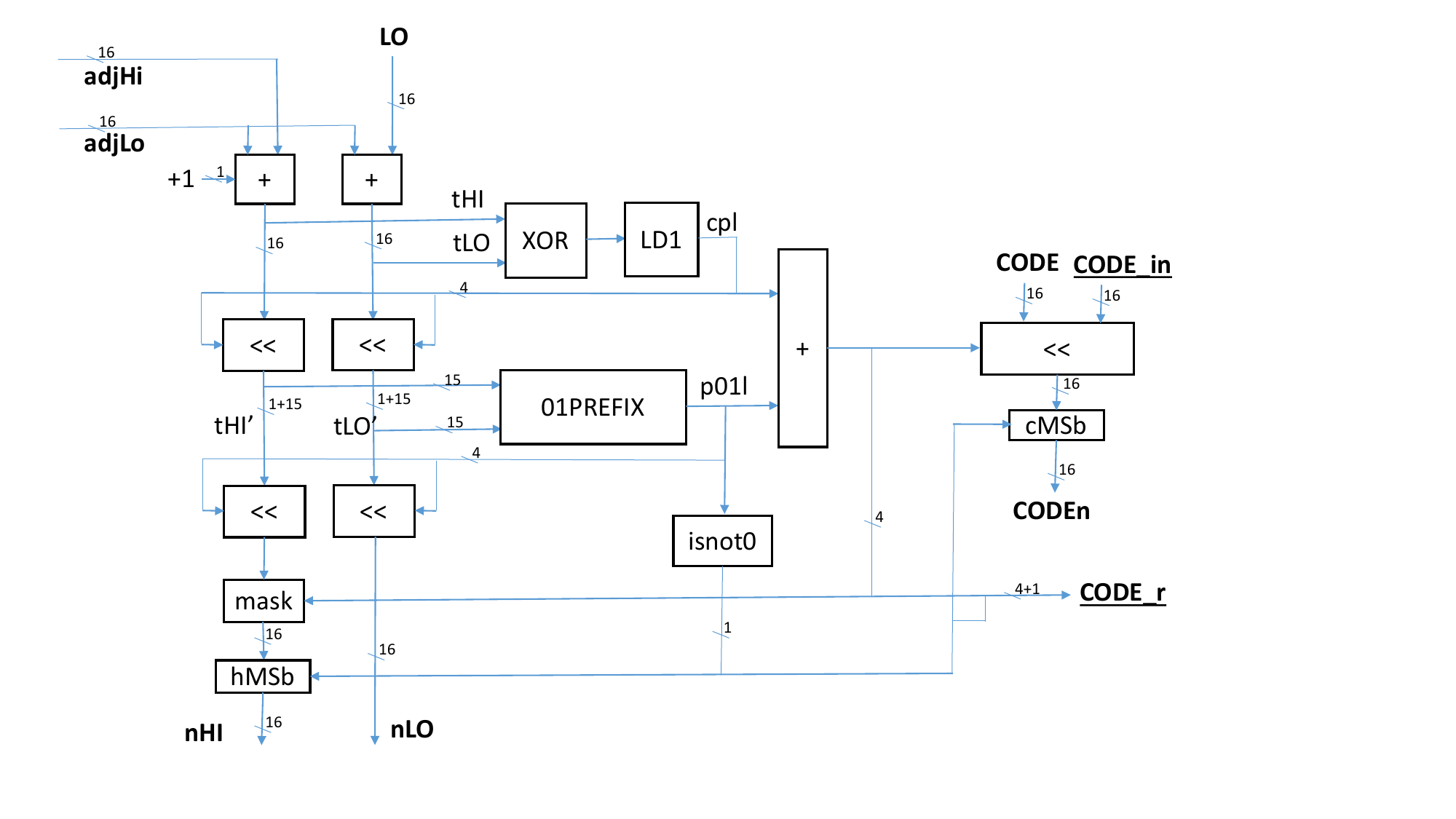}
\caption{HI/LO/CODE Adj}
\label{fig:dec:hilocode}
\end{subfigure}
\begin{subfigure}[b]{.25\textwidth}
\end{subfigure}
\caption{\OURLCORE\ Decoder: Greyed blocks are registers all else is combinatorial logic.}
\label{fig:dec:over}
\end{figure}

\subsection{Decoder Implementation}
\label{sec:impl:dec}
Figure~\ref{fig:dec:over} shows the \OURL decoder which is very similar to the encoder albeit a bit simpler. The decoder has the same initialization ports as the encoder (instead of a ``SYMBOL Lookup'' unit the decoder has a ``SYMBOL Gen'' unit -- however, internally these two units have an identical symbol table).  During decoding, it accepts two input streams CODE\_IN and OFS\_in respectively for reading bits from the encoded symbol and the offset streams. 

After initialization of the internal range, base, offset tables the decoder can start accepting the encoded symbol stream via CODE\_IN. At each step/cycle, the decoder reads in a number of bits and notifies whichever unit supplies them of their number via the 5b CODE\_r output. OFS\_in is similarly an $8b$ input that the decoder uses to read in offset bits (most significant bit first). How many it needs to read is given by the output OFS\_r which is $3b$ plus a $1b$ enable.

Figure~\ref{fig:dec:org} shows the internal organization of the encoder. There are four blocks: ``SYMBOL Gen'', ``PCNT Table'', ``Hi/Lo/CODE Adj'', and ``OFS Adj''. Two 16b registers HI and LO maintain the current range. These are initialized to 0xFFFF and 0x0000 respectively. The $16b$ CODE register reads in the encoded symbol stream, and the $8b$ OFS register is used to read in the offset stream.

Decoding a symbol is done by processing the current range boundary values HI and LO and CODE\_IN in ``PCNT\_Table'' (Figure~\ref{fig:dec:pcnt}). This is done by detecting which (scaled according to the HI and LO) probability count range the current CODE\_IN falls within. The results in identifying a symbol row index in SYMi and two adjusted range boundary values adjHI and adjLO. SYMi is used by ``SYMBOL Gen'' (Figure~\ref{fig:dec:symgen}) to produce the original value consuming the corresponding offset from the offset stream. The decoded value appears on the OUT port.
The ``HI/LO/CODE block (Figure~\ref{fig:dec:hilocode}) produces the new range boundary values, nHI and nLO, adjusts the CODE value, and potentially consumes bits from the encoded symbol stream. This mirrors the corresponding encoder process.

\noindent\textbf{Summary: }
The coding/decoding units prove simple, light-weight, and area- and energy-efficient. Their tables have very few entries (here 16) containing narrow fixed-point values. They are not actively updated. The rest of the logic is combinatorial implementing narrow fixed-point operations (the most expensive being a 16b x 10b multiplication out of which we don't need the lower 10b). The only updates done are to the HI, LO, CODE, and OFS registers which are short (16b).

\subsection{Increasing Throughput}

Processing neural networks places significant bandwidth demands on memory. The implementation described thus far can process only a single value per cycle. To meet the bandwidth demands of these workloads it is necessary to process multiple values per unit of time. This section describes two such methods: pipelining and replication.

\subsubsection{Pipelining the Encoder and Decoder}\label{sec:pipelining}

The implementation described encodes and decodes one value at a time. The implementation can be pipelined to increase the operating frequency, and to decrease the number of units needed to achieved a desired throughput hence improving area costs and energy efficiency. For this purpose, the input tensor is partitioned into a number of subtensors each of which are to be encoded and decoded as independent streams. Some state elements of the encoder and decoder unit need to be replicated to support the pipelined processing of multiple such streams. All the streams can use the same probability count table, replicating the PCNT (HiCnt[i]) and the symbol generation and lookup tables (base[i] and ob[i] fields) is not necessary. However, a separate set of the OFS, CODE, HI, and LO registers is needed per stream. Partitioning the encoder and the decoder into pipeline stages can be done in multiple ways. First the PCNT lookup can be partitioned into a number of stages where in each only a range of table entries are considered. For example, the table lookup can be partitioned into two stages, where the first considers the first half of entries and the second stage considers the second half. If the first stage results in a match, the second stage can be optionally power gated to reduce energy. Partitioning requires the introduction of temporary registers to hold the results produced at each stage. Adjusting the Hi/Lo/CODE can be another stage for the decoder, and generating the Hi/Lo/CODE can be another stage for the encoder. Similarly, the offset generation and offset extraction can be other stages operating.

\subsubsection{Replication}\label{sec:replication}

Pipelining increases throughput since it reduces the amount of processing that occurs per cycle thus permitting a higher operating frequency. Further increases in throughput can be had through replication of the encoder and decoder. This requires further splitting the input tensor into several subtensors whose stream are to be encoded and decoded independently. Such splitting is desirable for other reasons as it is often used to take advantage of locality in the on-chip memory hierarchy. The number of units needed depends on the target application and the number of processing cores. 

\section{\OURLCORE\ Table Generation}
\label{sec:profiler}
A method is proposed to generate the probability count table contents, such as those shown in Table~\ref{tbl:symt:ex}. Typically, for each layer of a model, the method is invoked twice to generate two separate tables, one for the activations and another for the weights. Since weights do not change a single pass over the weights is sufficient. Activations however depend on the input. Profiling proved effective since the overall distribution of activation values when viewed at the layer level does not change much with the input. 



Informally, the table generation method initializes the table with values corresponding to a uniform value distribution and estimates the resulting footprint, that is the number of bits that would be needed to store the input tensor. The method used to estimate the footprint is described later on. It then iteratively adjusts the table entries attempting to find a better assignment. This process considers all possible configurations allowed by the search algorithm as long as this results in a reduction in footprint that is above some threshold. The preferred threshold is set to 1\%. The method used is shown in Listing~\ref{lst:search}. 

Without loss of generality this description assumes that the inputs are 8b values, however, the same process can be applied to input of any bit length.  The process first inspects the input tensor and creates a histogram with $2^{8}$ buckets where each bucket $h(i)$ represents the number of times the corresponding value $i$ appears in the tensor. This is not shown in the listing. The process then starts by invoking the findPT() routine which accepts the histogram as input. The first step is to initialize the probability count table (PT) with a configuration where the range of possible values [0, $2^{8}-1$] is equally split among the table entries (line 38 in Listing~\ref{lst:search}). Function search() searches through candidate configurations and returns the best PT it found and its corresponding footprint. Line 43 decides whether the method will try to search for an even better configuration. This is done by testing the ratio of the newly found best footprint (newsize) over the previous best footprint (size). As long the improvement was above our 1\% threshold, the method will proceed to another round. Search uses recursion whose depth is controlled by the depth parameter. As long as depth is less than a maximum allowed value DEPTH\_MAX search() is allowed to call itself. A maximum depth of 2 was sufficient. The parameter ``around'' identifies the PT entry indexes would search() try to adjust. If ``around'' is negative then search() is allowed to adjust all entries. The only case this occurs is when findPT() invokes search(). Otherwise, ``around'' is an index itself and search() will try to adjust only entries whose distance from ``around'' is at most within some threshold either below or above. A distance threshold of 1 proved sufficient. Calls to encoded\_size() are used to estimate the compression ratio possible given the current table configuration. This is done by calculating the entropy of each range.

\lstdefinestyle{mystyle}{
    numberstyle=\footnotesize,
    basicstyle=\ttfamily\footnotesize,
    breakatwhitespace=false,         
    breaklines=true,                 
    captionpos=b,                    
    keepspaces=true,                 
    numbers=left,                    
    showspaces=false,                
    showstringspaces=false,
    showtabs=false,                  
    tabsize=2,xleftmargin=10em,framexleftmargin=5em
}
\lstset{style=mystyle}

\begin{figure}
\centering\ttfamily
\begin{lstlisting}[basicstyle=\tiny, mathescape=true, caption=Probability Count Table Generation, label=lst:search ]

N: number of probability count table entries
PT: Probability count table, array [1...N]
histogram: how many times each input value appears, array [0..VALUE_MAX]
depth: integer 
around: integer 1...N
DEPTH_MAX: integer, default 2
THRESHOLD: real, default 0.99

search(histogram, PT, minsize, depth, around)
  tryPT = PT
  for i=1 to N
    if around>=1 and $|$i - around$|$!=1: continue
    save=tryPT[i].$v_{min}$
    repeat   
        if i==1: $pv_{min}$=0
        else:    $pv_{min}$=tryPT[i-1].$v_{min}$
        if tryPT[i].$v_{min}$==$pv_{min}$: break
        tryPT[i].$v_{min}$--
        if depth > DEPTH_MAX:
           PT, minsize = search(histogram, tryPT, minsize, depth + 1, i)
        else:
           trysize = encoded_size(histogram, tryPT)
           if try_size<minsize: PT=tryPT, minsize=trysize
    tryPT[i].$v_{min}$=save
    repeat   
        if i==N: $nv_{min}$=VALUE_MAX
        else:    $nv_{min}$=tryPT[i+1].$v_{min}$
        if tryPT[i].$v_{min}$==$pv_{min}$: break
        tryPT[i].$v_{min}$++
        if depth > DEPTH_MAX:
           PT, best_size = search(histogram, tryPT, minsize, depth + 1, i)
        else:
           try_size = encoded_size(histogram, tryPT)
           if try_size<minsize: PT=tryPT, minsize=trysize
    return PT, minsize
    
findPT(histogram)
  initialize PT to uniform distribution
  repeat
    size=encoded_size(histogram, PT)
    PT, newsize = search(histogram, PT, size, 1, -1)
    if newsize/size>=THRESHOLD: break
  return PT
    
           
\end{lstlisting}
\end{figure}

\noindent\textit{Generating the Probability Counts: }
After the $v_{min}$ values are decided, they are used to generate probability counts. The probability counts range of $[0...2^m]$, where $m$ a design parameter (we use $m=10$), is partitioned proportional to the frequency of the values in each range given the $v_{min}$ configuration. 

\noindent\textit{Final Adjustment for Activations: }
When using profiling on activations it is possible that some values will not appear for the given set of inputs. However, just because a certain value did not appear when processing a limited set of input this is no guarantee that these values will never appear. A post processing step adjusts the probability count table by ``stealing'' a single count from another non-zero entry for each zero entry.

\section{Evaluation}\label{sec:eval}
\noindent\textbf{Compression Methods: }We compare \OURLCORE\ against: 1)~Baseline: that does not apply any compression to the off-chip traffic. 2)~RLE and RLEZ: run length encoding techniques that encode values as tuples of \textit{(value, distance)} where \textit{distance} is the number of similar values, or zeros, until the next non-similar, or non-zero, value for RLE and RLEZ respectively. The distance is limited to be up to 15 for a maximum overhead of 4-bit per tuple. 3)~ShapeShifter~\cite{Lascorz:2019:SEF:3352460.3358295}: a compression technique that, rather than using a fixed bit-width per value, groups the values in a predetermined group size \textit{G} and dynamically detects the minimal precision \textit{P} needed to represent the values in the group based on the actual range of values in the group. Then the group is represented with ${(G \times P + \log_2 P_{max})}$ bits where ${P_{max}}$ is the max precision supported per value. We evaluate a variant of ShapeShifter that is optimized for 8bit quantized models. 

\sloppy
\noindent\textbf{DNN models: } We evaluate the effectiveness of \OURL over a set of (quantized to int 8 unless otherwise noted) DNN models spanning a wide range of applications including image classification, object detection, segmentation, and natural language processing. The models we evaluate are listed in Table~\ref{tbl:models}. The models were obtained directly from the respective sources and are used unmodified.  Pruned versions of Alexnet and Googlenet allow us to study how \OURL interacts with pruned models~\cite{cvpr_2017_yang_energy}. From the Torchvision repository we use those models that have been pre-quantized to 8b fixed point~\cite{torchvision}. From the IntelAI repository we similarly use only those models whose weights are quantized to 8b fixed-point~\cite{intelai}.
ResNet18-PACT was quantized to 4b except for the first and last layers which remain in 8b using the PACT method of Choi et al.~\cite{choi2018pact}. ResNet18-Q is pruned with the modified training method of BitPruning to arrive at per layer fixed-point precisions that never exceed 8b~\cite{nikoli2020bitpruning}. The ``per-layer'' quantized models were quantized further with the profiling-based quantization method of Nikolić et al.~\cite{8695654} which further trims precisions per layer.

\noindent\textbf{Trace Collection: } We run inference with the models out-of-the-box as released by their original authors on an NVIDIA GeForce RTX 3090 GPU. We use PyTorch and TensorFlow layer hooks that trigger upon executing each layer to dump its input weights and activations into numpy files. Since model parameters do not change while inference, a single trace dump for weights is enough per layer where the same trace is used to derive the probability count table and then compressed using the table. However, input activations are changing dynamically with every input image to the model. Thus, up to 9 input activation samples per layer are used to generate the probability tables. The IntelAI models, as provided, use floating-point activations. Hence we limit attention only to their weights.

\noindent\textbf{Hardware and Energy Modeling}: We attach 64 \OURL encoder/decoder engines to a dual-channel 8GB DDR4-3200 off-chip memory interface for evaluation. The energy consumption of the baseline off-chip memory accesses was modeled using Micron's DRAM power model~\cite{micron} along with the uncompressed data volume. For \OURL, the data was compressed using the profiling-based probability count tables then passed through the same DRAM power model while taking the overhead power consumption of \OURL engines into account. To model the area and power consumption of \OURL, the compression/decompression engines were implemented in Verilog, synthesized via the Synopsys Design Compiler and layout was produced via Cadence Innovus and for a 65nm TSMC\ technology which is the best that is available to us due to licensing restrictions. The power consumption overhead of the encoding/decoding engines was estimated by capturing circuit activity using Mentor Graphics' ModelSim which was then passed on to Innovus for post-layout simulation.

To compare with ShapeShifter we follow the same modeling flow where we model compression/decompression engines using the same technology node and design flow. We use a configuration optimized for 8bit quantized models with a single Level-1 unit along with eight Level-2 units are used per off-chip DRAM channel~\cite{Lascorz:2019:SEF:3352460.3358295}.

\begin{table*}[h!]
        \centering
        \caption{Neural network models studied.}
        \label{tbl:models}
\vspace{-5pt}
\rmfamily\footnotesize
\begin{tabular}{cccccc}
\textbf{Network} & \textbf{Dataset} & \textbf{Application} & \textbf{Data Type} & \textbf{Quantizer}  \\
\hline
GoogLeNet~\cite{googlenet} & ImageNet~\cite{imagenet} & Classification  & int8 & Torchvision\\
Inception v3~\cite{inceptionv3} & ImageNet & Classification  & int8 & Torchvision\\
Mobilenet v2~\cite{MobileNetV2} & ImageNet & Classification  & int8 & Torchvision\\
Mobilenet v3~\cite{MobileNetV3} & ImageNet& Classification  & int8 & Torchvision\\
Resnet18~\cite{resnet} & ImageNet & Classification  & int8 & Torchvision\\
Resnet50~\cite{resnet} & ImageNet & Classification  & int8 & Torchvision\\
Resnext101~\cite{resnext} & ImageNet & Classification  & int8 & Torchvision\\
Shufflenet v2~\cite{shufflenetv2} & ImageNet & Classification  & int8 & Torchvision\\
Inception v4~\cite{Inceptionv4} & ImageNet & Classification  & int8 & IntelAI\\
Mobilenet v1~\cite{MobileNets} & ImageNet & Classification  & int8 & IntelAI\\
Resnet101~\cite{resnet} & ImageNet & Classification  & int8 & IntelAI\\
R-FCN Resnet101~\cite{MobileNetV3} & COCO~\cite{lin2014microsoft}& Object Detection & int8 & IntelAI\\
SSD-Resnet34~\cite{SSD_Resnet} & COCO & Object Detection  & int8 & IntelAI\\
Wide \& Deep~\cite{wide_deep} & Kaggle Display Advertising Dataset~\cite{kaggleAdDataset} & Recommendation  & int8 & IntelAI\\
Q8BERT~\cite{BERT,izsak_peter_2018_1477518} & MRPC~\cite{GLUE} & NLP & int8 & IntelLabs Distiller\\
NCF~\cite{NCF,Distiller} & ml-20m~\cite{ml20m} & Recommendation & int8 &  IntelLabs Distiller+Per-Layer\\
ResNet18-PACT~\cite{choi2018pact}  & ImageNet   & Classification  & int4/int8 & IntelLabs Distiller+Per Layer\\
SSD-MobileNet~\cite{Liu_2016,reddi2019mlperf}  & COCO~\cite{lin2014microsoft} & Object Detection  & int8 &  MLPerf+Per Layer\\
MobileNet~\cite{MobileNets,reddi2019mlperf} & ImageNet & Classification  & int8 &  MLPerf\\
bilstm~\cite{bidirectionalLSTM}  & Flickr8k~\cite{flickr8k}    & Captioning  & int8 &  per-layer\\
SegNet~\cite{badrinarayanan2015segnet} & CamVid~\cite{BrostowFC:PRL2008} & Segmentation & int8 &  per layer\\
ResNet18-Q~\cite{DBLP:journals/corr/HeZRS15,nikoli2020bitpruning} & ImageNet & Classification  & int8  & per-layer\\
AlexNet-Eyeriss~\cite{cvpr_2017_yang_energy} & ImageNet & Classification  & int8/Pruned  & per-layer\\
GoogLeNet-Eyeriss~\cite{cvpr_2017_yang_energy} & ImageNet & Classification  & int8/Pruned  & per layer\\
\hline
\end{tabular}
\vspace{-10pt}
\end{table*}

\subsection{Change in Memory Traffic}

\begin{figure*} 
\centering
\begin{subfigure}[b]{.49\linewidth}
\centering
\includegraphics[width=\linewidth]{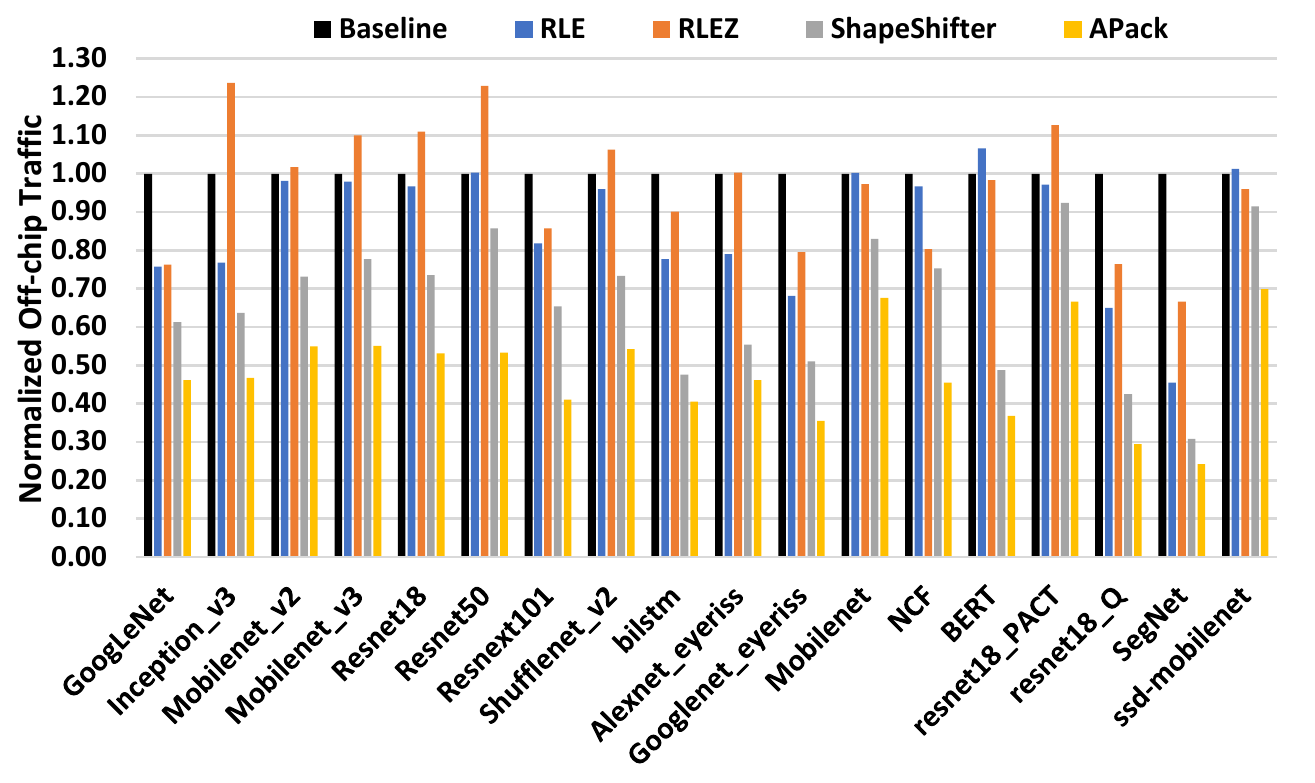}
\caption{Activations}
\label{fig:activ_compression}
\end{subfigure}
\begin{subfigure}[b]{.49\linewidth}
\centering
\includegraphics[width=\linewidth]{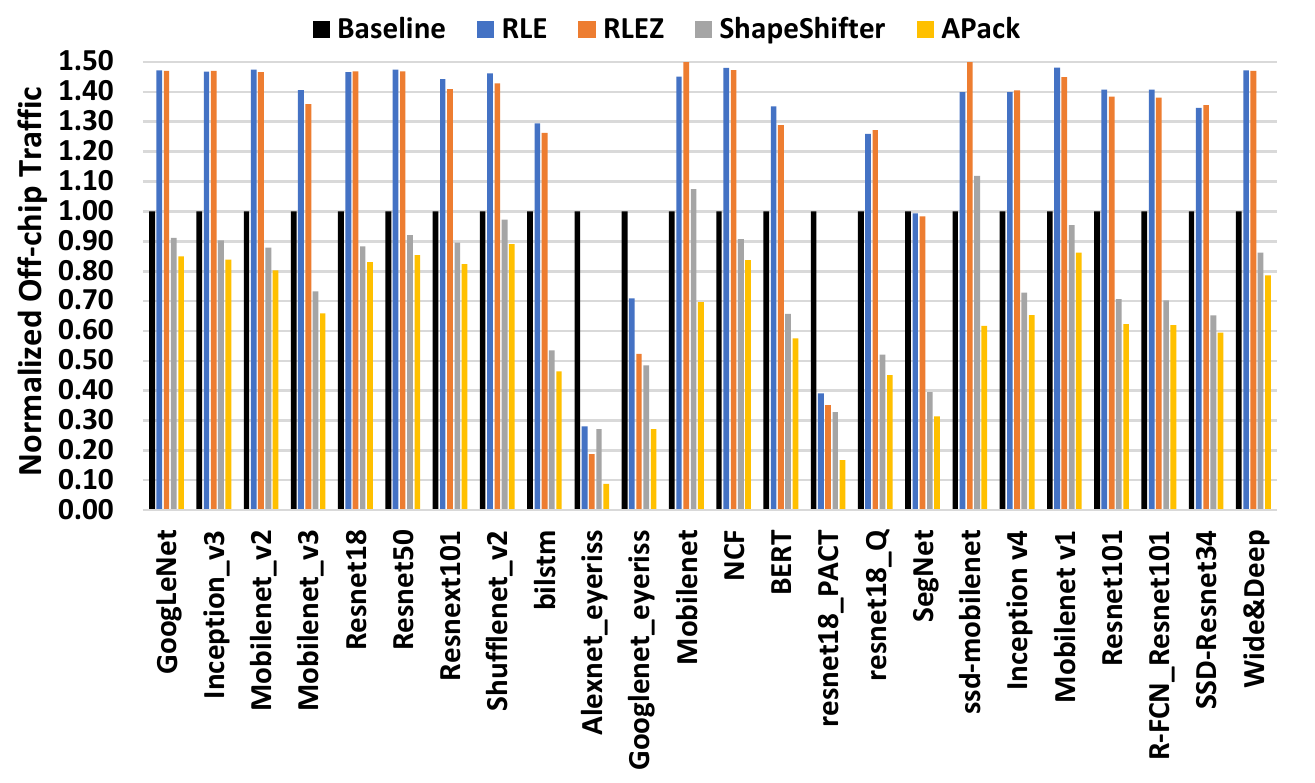}
\caption{Weights}
\label{fig:weights_compression}
\end{subfigure}
\caption{Normalized off-chip traffic}
\vspace{-10pt}
\end{figure*}

This section reports the change in off-chip memory traffic relative to the baseline (no compression) for \OURL, Shapeshifter (SS)~\cite{Lascorz:2019:SEF:3352460.3358295}, Run-Length Encoding (RLE), and Run-Length Encoding for Zeros (RLEZ)~\cite{isscc_2016_chen_eyeriss, EIEISCA16, cambricon:2016}. Figures~\ref{fig:weights_compression} and~\ref{fig:activ_compression} report the relative reduction in off-chip traffic for weights and activations respectively. The results demonstrate that \OURL is robust in that it always reduces traffic, it outperforms the other methods, and that the reductions in traffic depend on the quantization method used. Generally, the reduction is higher for activations than for weights except for when the models are pruned. The rest of this section discusses these results in more detail as the quantization method used by each model affects value distributions during inference.


\noindent\textbf{Torchvision models}:
For model parameters, \OURL reduces traffic to as much 0.65 of the baseline for MobileNet v3 and to as little as 0.88 for ShuffleNet v2. Much higher reductions are observed with \OURL for activations: the highest reduction is to 0.41 of the baseline for for ResNext101 and the lowest is 0.55 for MobileNet v3. There are two reasons for the better compression of activations. First, activations show high sparsity,  a well known feature of neural networks that use the ReLU activation function which clips most non-firing neurons to zero.  

Inspecting the weight distribution we found that while most values do cluster near zero or near the maximum, many of the intermediate values are present. When compared to the weight distribution of the other models that were quantized with different methods, this suggests that the lower bits tend to be noisy, a tell-tale sign that the quantization method used by TorchVision uses the full value range regardless of whether it is needed. A similar phenomenon was observed for the linear quantization used in earlier versions of Tensorflow~\cite{DBLP:journals/corr/DelmasSJM17}.

Shapeshifter and the run-length-based method have a harder time coping with this noisy distribution. Shapeshifter manages to reduce traffic but less compared to \OURL. RLE and RLEZ result in increasing traffic for weights as repetition of values, be it zeros or otherwise, is rare. They do offer some reduction in traffic for activations. As expected, \OURL outperforms Shapeshifter and the other methods for weights and more so for activations. Shapeshifter groups values (we used a group of 8 values as in the original work which we verified that works best for these models too) and uses a container that is as larger as necessary to accommodate the largest magnitude value within the group. Encoding efficiency is lost for all other values having a lower magnitude. In contrast, \OURL treats each value independently and effectively uses as many bits, even a fractional number, to encode it as required by the value's frequency of appearance.

\noindent\textbf{IntelAI models}:
The IntelAI quantized models demonstrate that having a noisy weight distribution is not inherently necessary for weights. In general, compared to the TorchVision models, IntelAI's quantization methods results in more skewed distributions for weights, a property that \OURL rewards resulting in pronounced compression benefits for weights compared to the TorchVision models. \OURL reduces traffic to as little as 0.59 of the baseline for SSD-Resnet34 and in the worst case to 0.86 for MobileNet v1 with most of the models observing a reduction to 0.6. Compared to ShapeShifter \OURL reduces traffic by 11\% on average.  The runlenth-based methods still fail to improve memory footprints for the weights as these models are also not pruned.

\noindent\textbf{Remaining models}:
For the remaining models and relative to ShapeShifter, \OURLCORE\ compresses activations and weights by $1.34\times$ and $1.51\times$ respectively. The average reduction is higher for weights due to the highly sparse and skewed distribution of the model parameters, especially for AlexNet\_Eyeriss, Googlenet\_Eyeriss, and ResNet18\_PACT. The results also demonstrate that \OURL benefits all quantization methods. For example, it reduces traffic even for ResNet18\_PACT taking advantage of the skewed distribution of the 4b weights and much more than ShapeShifter does. The run-length-based methods have their best showing for the pruned models. Even for them, however, \OURL is nearly twice as effective.

\subsection{Area, Power, and Energy}
This section reports the energy savings that \OURL achieves with the compressed off-chip traffic to/from DRAM. We model the power consumption of a dual-channel 16GB DDR4-3200  off-chip memory via Micron's DRAM power model~\cite{micron} and use 64 \OURL compressors/decompressors.

\begin{figure*} 
\centering
\begin{subfigure}[b]{.49\linewidth}
\centering
\includegraphics[width=\linewidth]{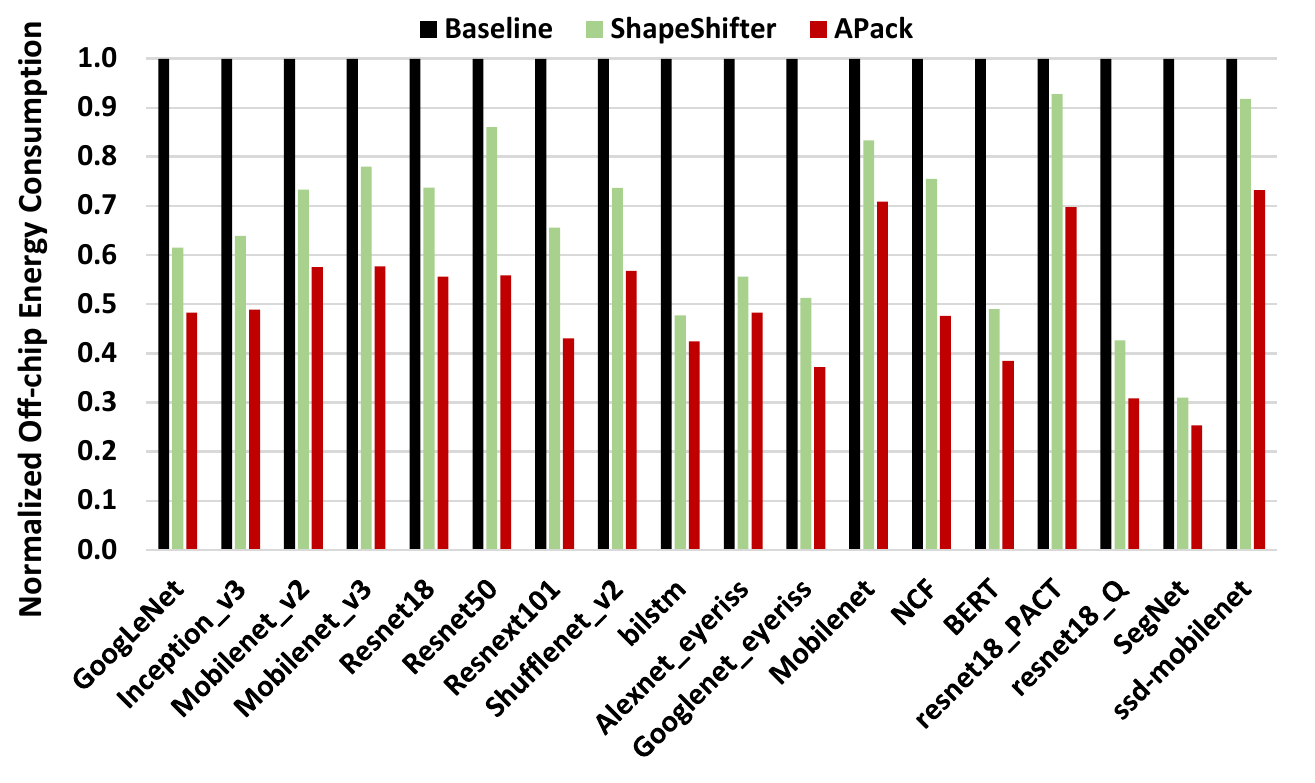}
\caption{Activations}
\label{fig:activ_energy}
\end{subfigure}
\begin{subfigure}[b]{.49\linewidth}
\centering
\includegraphics[width=\linewidth]{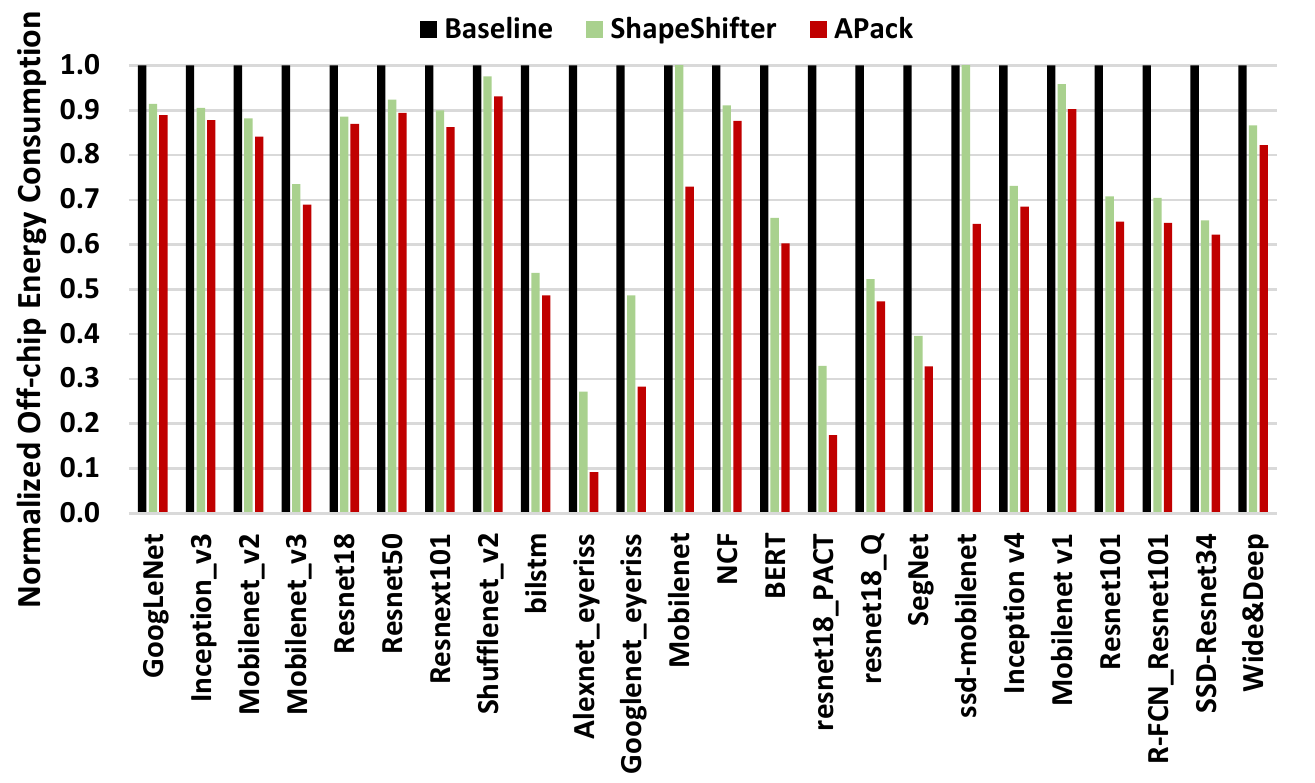}
\caption{Weights}
\label{fig:weights_energy}
\end{subfigure}
\caption{Normalized off-chip energy consumption}
\label{fig:energy}
\vspace{-10pt}
\end{figure*}

Our layout-based measurements are that the 64 compression/decompression engines of \OURLCORE\ need a total area of $1.14\ mm^2$ and consume a total power of $179.2\ mW$. This power consumption constitutes an $4.7\%$ overhead vs. the power consumed by a dual-channel DDR4-3200 memory system when operating at 90\% of its peak bandwidth.

Figure~\ref{fig:energy} shows the energy consumed by off-chip memory transfers normalized to the baseline that implements no compression technique. The estimates shown assume that the weights and input activations of each layer are read only \textit{once} from off-chip~\cite{kevinmemory}. This is the typical case for real-time ``edge'' inference accelerators where the whole DNN model cannot fit in on-chip memory and, thus, the parameters of each layer should be read from off-chip for each single input image passing through the model. 

The energy savings vary across models according to the compression ratio achievable. For instance, for pruned models such as AlexNet-Eyeriss and GoogLeNet-Eyeriss where most of the model weights are zeros, \OURL capitalizes on the heavily skewed distribution and saves 91\% and 72\% of the off-chip access energy respectively. For models with less skewed distributions, such as NCF \OURL achieves lower compression ratio ($1.2\times$) and but still saves 13\% of the off-chip energy consumption. The activation distributions are more heavily skewed. For example, the compression ratio for the activations for NCF is $2.2\times$ leading to energy savings of 53\%. Generally, the energy savings results track the achievable compression ratios thanks to the light power consumption overhead of the \OURLCORE\ compressor/decompressor units.

Although ShapeShifter's hardware and power costs are lower than \OURL's, the higher compression with \OURL and given that off-chip accesses are so expensive in comparison to the compressors/decompressors result in superior overall energy with \OURL. 

\subsection{Overall Performance and Energy Efficiency}

We measure end-to-end energy efficiency and speedup when \OURLCORE\ is integrated with a Tensorcore-based accelerator. The accelerator is configured as shown in Table~\ref{tbl:arch_config_TCs}. We evaluate both \OURLCORE\ and ShapeShifter for off-chip compression and compare to the baseline accelerator that does not apply any off-chip compression technique. We obtained the Shapeshifter simulator from the original authors and modify its memory and processing energy modeling for consistent modeling of all configurations. We limit attention to the only those model traces that are compatible with this simulator. While overall \OURL achieves higher compression rates than ShapeShifter, for these models, as we have seen the advantage of \OURL is smaller. Accordingly, the performance and energy advantage of \OURL over ShapeShifter would have been higher if we were able to measure them across all previously studied models.

\begin{figure} 
\centering
\includegraphics[width=.9\linewidth]{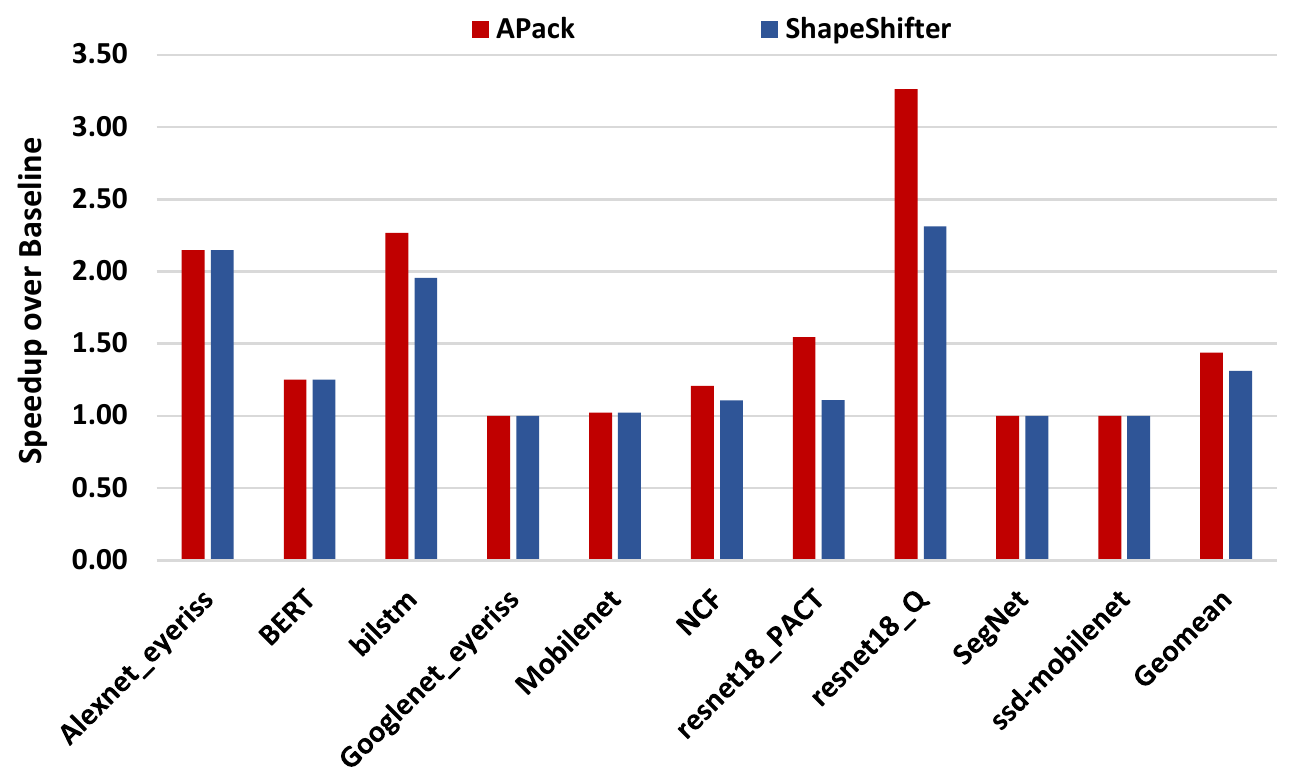}
\caption{Overall speedup.}
\label{fig:TC_speedup}
\vspace{-10pt}
\end{figure}

Figure~\ref{fig:TC_speedup} shows the speedup achieved when \OURLCORE\ enhances the baseline accelerator. On average, \OURLCORE\ speeds up the execution by $1.44\times$ while ShapeShifter achieves a $1.3\times$ speedup over the baseline. Both methods avoid stalls for off-chip transfers and enabling much better use of the accelerator's compute units. This is especially true for models that tend to be memory-bound with low compute per byte ratio. For all these models, \OURL achieves better performance than ShapeShifter. For some, however, the execution time advantage of \OURL is minimal. These are models such as BERT and the pruned Alexnet and GoogleNet, that become completely compute compute bound. However, as we will see when we consider overall energy, \OURL has a significant edge over ShapeShifter for all models as it reduces off-chip traffic more. 

\begin{figure} 
\centering
\includegraphics[width=.9\linewidth]{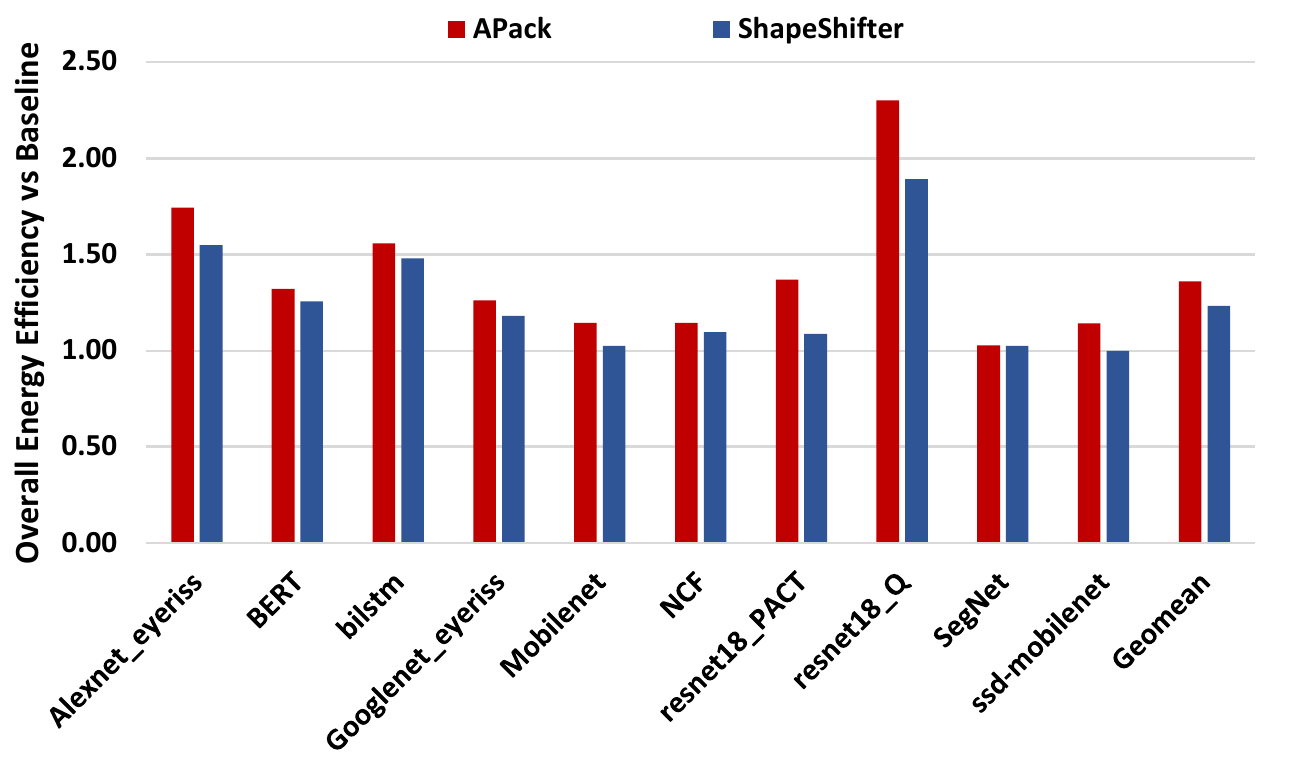}
\caption{Overall energy efficiency.}
\label{fig:TC_EE}
\vspace{-10pt}
\end{figure}

Figure~\ref{fig:TC_EE} shows that \OURL boosts the energy efficiency over the baseline accelerator for all the experimented models. The improvement varies per model according to the relative importance of the off-chip transfer energy vs. the on-chip compute energy. The higher the fraction of overall energy due to off-chop accesses the higher the potential and the benefits seen from \OURL. On average over all models, \OURLCORE\ boosts energy efficiency by $1.37\times$ outperforming ShapeShifter which achieves a $1.23\times$ energy efficiency over the baseline.

\begin{table}
\centering
{
\scriptsize
\begin{tabular}{|l|l|l|l|}
\hline
\multicolumn{4}{|c|}{\textbf{TensorCore-based Accelerator}} \\\hline 
\# of TCs        & 64   &  Activations Buff. & $256KB{\times}16$ Banks \\ \hline
TC core   & $4\times4$ PEs & Weights Buff& $256KB{\times}16$ Banks \\ \hline
PE MACs/Cycle & 4  &  Output Buff. & $256KB{\times}16$ Banks \\ \hline
Tech Node    & 65nm & Frequency    & 1 GHz \\\hline
     \multicolumn{2}{|c|}{Off-Chip Memory}
     & \multicolumn{2}{c|}{8GB 2-channel DDR4-3200} \\ \hline
     \multicolumn{2}{|c|}{Peak TOPS}
     & \multicolumn{2}{c|}{8.2 TOPS (int8)} \\ \hline
\end{tabular}
}
\caption{{Tensorcore-based accelerator configuration.}}
\label{tbl:arch_config_TCs}
                \vspace{-10pt}
\end{table}

\section{Related Work}\label{sec:related}

The most closely related work to \OURL are compression methods that target deep learning inference. As commented in the introduction, these methods target specific value patterns such as zeros, or prefixes of 0s and 1s~\cite{han_deep_2015-1,Lascorz:2019:SEF:3352460.3358295,DBLP:conf/hpca/RhuOCPKK18,cavigelli2018bitPlaneCompr,BOVEDA}. Bit-interleaving across several values~\cite{cavigelli2018bitPlaneCompr} and precision trimming quantization~\cite{judd_proteus_wapco_2016,nikoli2020bitpruning} has been used to amplify how often such patterns appear. \OURL differs in that it is not designed to target any specific value pattern whether that is for whole values or bits. It is designed to target \textit{any} non-uniformity in the value distribution and for that could reward methods that amplify them.

The introduction has also commented on several other approaches that can boost model efficiency. As long the resulting model exhibits skewed value distribtions, \OURL will complement such methods.

\section{Conclusion}
\label{sec:conclude}
\OURL is a simple to implement and effective off-chip compression method for neural networks that is plug-in compatible with many accelerators. We have demonstrated that it reduces off-chip traffic which significantly reduces energy consumed by expensive off-chip memory transfers. In addition, \OURL can accelerate memory-bound models by reducing the stalls related to off-chip data accesses. \OURL greatly boosts overall energy efficiency and proves superior to prior lossless memory compression techniques for deep learning.  


\section*{Acknowledgements}
This work was supported by an NSERC Discovery Grant and the NSERC COHESA Strategic Research Network. The University of Toronto maintains ownership of the technologies described.

\bibliographystyle{IEEEtranS}
\bibliography{refs}

\end{document}